\documentclass[runningheads,a4paper]{llncs}

\usepackage[top=2.0cm, bottom=2.0cm, left=2.5cm, right=2.5cm]{geometry}
\usepackage{amsmath}
\usepackage{amssymb}
\usepackage{graphicx}
\usepackage{booktabs}
\usepackage[dvipsnames,table]{xcolor}
\usepackage{url}
\usepackage{microtype}
\usepackage[T1]{fontenc}
\usepackage{algorithm}
\usepackage{algorithmicx}
\usepackage{algpseudocode}
\usepackage{siunitx}
\usepackage{multirow}
\usepackage{tabularx}
\usepackage{colortbl}
\usepackage[most]{tcolorbox}
\usepackage{longtable}
\usepackage{caption}
\usepackage{makecell}
\usepackage{ragged2e}
\newcolumntype{L}[1]{>{\RaggedRight\arraybackslash}p{#1}}
\usepackage{threeparttablex}
\usepackage{subcaption}
\usepackage{array}
\usepackage{titlesec}
\usepackage{lineno}
\usepackage[
    colorlinks=true,
    citecolor=MidnightBlue,
    linkcolor=red,
    urlcolor=magenta
]{hyperref}
\newtcolorbox{highlighted}{
  colback=black!5,
  colframe=yellow!50!black,
  boxrule=0pt,
  sharp corners,
  breakable,
  before skip=2pt, after skip=2pt,
  left=0pt, right=0pt, top=1pt, bottom=1pt,
  boxsep=2pt,
  parbox=false,
}
\usepackage{cleveref}

\makeatletter

\renewcommand{\thesubsection}{\arabic{subsection}}
\renewcommand{\thesubsubsection}{\thesubsection.\arabic{subsubsection}}

\@addtoreset{subsection}{section}
\@addtoreset{subsubsection}{subsection}

\titleformat{\subsection}[hang]
    {\normalfont\large\bfseries}
    {\thesubsection}
    {1em}
    {}
    [\vspace{0.5ex}]

\titlespacing*{\subsubsection}{0pt}{12pt}{6pt}
\titleformat{\subsubsection}[hang]
    {\normalfont\normalsize\bfseries}
    {\thesubsubsection}
    {1em}
    {}
    [\vspace{0.5ex}]
\makeatother

\crefname{section}{Section}{Sections}
\Crefname{section}{Section}{Sections}
\crefname{subsection}{Subsection}{Subsections}
\Crefname{subsection}{Subsection}{Subsections}
\crefname{subsubsection}{Subsection}{Subsections}
\Crefname{subsubsection}{Subsection}{Subsections}

\title{CADS: A Comprehensive Anatomical Dataset and Segmentation for Whole-Body Anatomy in Computed Tomography}
\titlerunning{CADS}
\author{
    {\small Murong Xu\inst{1,*}}
    \and
    {\small Fernando Navarro\inst{1}}
    \and
    {\small Maksym Fritsak\inst{2,3}}
    \and
    {\small Ibrahim Ethem Hamamci\inst{1}}
    \and
    {\small Suprosanna Shit\inst{1}}
    \and \\
    {\small Bastian Wittmann\inst{1}}
    \and
    {\small Sezgin Er\inst{1,4}}
    \and
    {\small Sebastian M. Christ\inst{2}}
    \and
    {\small Ezequiel de la Rosa\inst{1}}
    \and
    {\small Julian Deseoe\inst{1,5}}
    \and
    {\small Robert Graf\inst{6,7}}
    \and \\
    {\small Hendrik M\"oller\inst{6,7}}
    \and
    {\small Anjany Sekuboyina\inst{1,6}}
    \and
    {\small Jan C. Peeken\inst{8,9,10}}
    \and
    {\small Sven S. Becker\inst{11,12}}
    \and
    {\small Giulia Baldini\inst{11,12}}
    \and \\
    {\small Johannes Haubold\inst{11,12}}
    \and
    {\small Felix Nensa\inst{11,12}}
    \and
    {\small Ren\'e Hosch\inst{11,12}}
    \and
    {\small Nikhil Mirajkar\inst{13,14}}
    \and
    {\small Saad Khalid\inst{14}}
    \and
    {\small Stefan Zachow\inst{15}}
    \and
    {\small Marc-Andr\'e Weber\inst{16}}
    \and
    {\small Georg Langs\inst{17}}
    \and
    {\small Jakob Wasserthal\inst{18}}
    \and
    {\small Mehmet Kemal Ozdemir\inst{19}}
    \and
    {\small Andrey Fedorov\inst{20}}
    \and \\
    {\small Ron Kikinis\inst{20}}
    \and
    {\small Stephanie Tanadini-Lang\inst{2}}
    \and
    {\small Jan S. Kirschke\inst{6}}
    \and
    {\small Stephanie E. Combs\inst{8,9,10}}
    \and \\
    {\small Tamaz Amiranashvili\inst{1,21,\dag}}
    \and
    {\small Bjoern Menze\inst{1,\dag}}
}
\authorrunning{Xu et al.}

\institute{
{\small{
$^{1}$ Department of Quantitative Biomedicine, University of Zurich, Zurich, Switzerland \\ \quad
$^{2}$ Department of Radiation Oncology, University Hospital and University of Zurich, Zurich, Switzerland \\ \quad
$^{3}$ Faculty of Medicine, University of Zurich, Zurich, Switzerland \\ \quad
$^{4}$ International School of Medicine, Istanbul Medipol University, Istanbul, Turkey \\ \quad
$^{5}$ Department of Neurology and Clinical Neuroscience Center, University Hospital Zurich and University of Zurich, Zurich, Switzerland \\ \quad
$^{6}$ Department of Diagnostic and Interventional Neuroradiology, School of Medicine and Health, Technical University of Munich, Munich, Germany \\ \quad
$^{7}$ Institut f\"ur KI und Informatik in der Medizin, Klinikum rechts der Isar, TUM School of Medicine and Health and School of Computation, Information and Technology, Munich, Germany \\ \quad
$^{8}$ Department of Radiation Oncology, TUM University Hospital Rechts der Isar, TUM School of Medicine and Health, Technical University of Munich, Munich, Germany \\ \quad
$^{9}$ Institute of Radiation Medicine (IRM), Helmholtz Zentrum München (HMGU), Munich, Germany \\ \quad
$^{10}$ German Consortium for Translational Cancer Research (DKTK), Partner Site Munich, Munich, Germany \\ \quad
$^{11}$ University Hospital Essen, Institute of Interventional and Diagnostic Radiology and Neuroradiology, Essen, Germany \\ \quad
$^{12}$ University Hospital Essen, Institute for Artificial Intelligence in Medicine, Essen, Germany \\ \quad
$^{13}$ Department of Radiology, Cambridge University Hospitals NHS Foundation Trust, Cambridge, UK \\ \quad
$^{14}$ Labelata GmbH, Zurich, Switzerland \\ \quad
$^{15}$ Visual and Data-Centric Computing, Zuse Institute Berlin (ZIB), Berlin, Germany \\ \quad
$^{16}$ Institute of Diagnostic and Interventional Radiology, Pediatric Radiology and Neuroradiology, University Medical Center Rostock, Rostock, Germany \\ \quad
$^{17}$ Department of Biomedical Imaging and Image-guided Therapy, Medical University of Vienna, Vienna, Austria \\ \quad
$^{18}$ Department of Radiology, University Hospital Basel, Basel, Switzerland \\ \quad
$^{19}$ School of Engineering and Natural Sciences, Artificial Intelligence Engineering Department, Istanbul Medipol University, Istanbul, Turkiye \\ \quad
$^{20}$ Brigham and Women’s Hospital, Boston, Massachusetts, USA \\ \quad
$^{21}$ School of Computation, Information and Technology, Technical University of Munich, Munich, Germany \\ \quad
}
}
~
\\
~
\\
\small{\dag}Tamaz Amiranashvili and Bjoern Menze jointly supervised this work.
\\
\small{*}Corresponding author: \email{murong.xu@uzh.ch}.
}

\begin{document}

\mainmatter
\maketitle
\setcounter{footnote}{0} 

\begin{abstract}
Volumetric computed tomography (CT) segmentation supports quantitative image analysis, but most AI-based methods remain organ-specific, with fragmented models, inconsistent label definitions, and heterogeneous training and evaluation protocols.
Here we present CADS, a unified data--model framework for scalable and generalizable whole-body CT segmentation.
We construct a SNOMED-CT-referenced anatomical label space harmonizing heterogeneous cohorts from over 100 imaging centers, yielding complete annotations of 167 structures across \num{22022} CT volumes and markedly expanding anatomical coverage and scale beyond existing resources.
To complete labels on sparsely annotated CTs, we combine self-training with shape-informed quality control based on neural implicit functions, curating high-quality pseudo-labels.
Trained on this resource, CADS shows robust whole-body segmentation across distribution shifts and improved performance relative to state-of-the-art approaches across 18 public datasets.
We further validate it in an independent, expert-reviewed real-world oncology cohort of \num{2864} CTs.
The curated dataset, pretrained models, and user-friendly software together provide an accessible, reproducible foundation for large-scale anatomical segmentation and computational anatomy research.
\end{abstract}
\section*{Introduction} \label{sec_01_introduction}

Computed tomography (CT) provides volumetric views of internal anatomy across diverse clinical settings.
Whole-body CT segmentation converts this information into structured anatomical representations, supporting downstream applications such as radiation-treatment planning~\cite{shi2022deep}, abnormality detection and triage~\cite{eyuboglu2021multi}, metastasis localization~\cite{vagenas2022decision}, and body-composition analysis~\cite{weston2019automated}.

Recent advances in segmentation have broadened the methodological landscape, spanning supervised~\cite{ronneberger2015u,milletari2016v,roth2015deeporgan}, prompt-based~\cite{ma2024segment,wong2024scribbleprompt,he2024vista3d,lin2026progidiff}, and in-context learning approaches~\cite{butoi2023universeg,ren2024medical}. 
As segmentation architectures mature, attention has shifted toward large-scale curated datasets, as strong baselines such as nnU-Net~\cite{isensee2021nnu} have shown that careful configuration and training can rival specialized network designs, making data quality, diversity, and scale increasingly important for further gains.
Yet, unlike natural-image corpora~\cite{dosovitskiy2020image,singh2023effectiveness,deng2009imagenet,krizhevsky2009learning}, medical datasets remain orders of magnitude smaller and sparsely annotated.
Across 40 public CT data sources included in this study, only 21\% of scans were accompanied by anatomical annotations, most of which covered only one or two structures (Table~\ref{table:dataset_collection}).

Large-scale efforts such as TotalSegmentator~\cite{totalsegmentator} have shown the feasibility of AI-assisted dataset creation and inspired subsequent methodological advances~\cite{diaz2024monai,plotka2024swin,ji2023continual}.
However, current resources remain constrained by
limited anatomical coverage, often restricted to selected body regions rather than whole-body anatomy~\cite{li2024abdomenatlas,liu2024cosst};
insufficient data diversity, which weakens generalization across scanners and populations~\cite{totalsegmentator,liu2024cosst,sundar2022fully,ji2023continual,jaus2023towards};
inconsistent annotation quality, as systematic labeling errors in complex structures can propagate through training~\cite{totalsegmentator};
and human-dependent workflows that constrain scalability and increase operational cost~\cite{sundar2022fully,li2024abdomenatlas}.
Further barriers include inadequate validation, where limited external evaluation undermines real-world robustness~\cite{liu2024cosst,sundar2022fully,ji2023continual,jaus2023towards};
restricted availability of models or datasets~\cite{liu2024cosst,ji2023continual}; and a lack of
accessible tools, which together hinder reproducibility, robustness assessment, and broader adoption.

To address these gaps, 
we develop CADS (Fig.~\ref{fig:results_whole_body_parts_visualization}) from a data-centric perspective, integrating heterogeneous CT cohorts into a unified anatomical label space and expanding sparse annotations through pseudo-labeling guided by shape-informed quality control. 
The resulting resource enables systematic investigation of how large-scale curated training data affect whole-body segmentation performance.
CADS therefore comprises two main components.

CADS-dataset: A large-scale collection of \num{22022} CT volumes with full annotations for 167 anatomical segmentation targets.
To our knowledge, it is among the largest publicly available whole-body CT resources, surpassing prior collections~\cite{totalsegmentator} in both scale (\num{18}$\times$ as many scans) and anatomical coverage (60\% more distinct targets).
Beyond scale, CADS provides a framework for foundation-scale data curation in medical image analysis. 
Our automated annotation pipeline unifies heterogeneous cohorts through pseudo-labeling, enabling scans with no or partial annotations to be incorporated into a common label space. 
We design a shape-informed quality-control mechanism based on neural implicit representations to systematically evaluate multi-class anatomical plausibility and select high-quality labels at scale.
Combined with a multi-expert training and label-assembly strategy, this reduces the influence of noisy pseudo-labels and supports learning from mixed supervision, including manual annotations and curated pseudo-labels.
The framework integrates CT data from 40 sources, including public archives (e.g., TCIA~\cite{tcia}), community challenges~\cite{grandchallenge,miccaichallenge}, and hospital cohorts, all harmonized under a unified anatomical ontology. 
We further contribute two new datasets comprising 484 head CTs and 586 triphasic contrast-enhanced abdominal CTs. 
Together, these data span more than 100 imaging facilities across 16 countries, capturing variations in clinical protocols and population characteristics. 
By coupling large-scale integration with automated quality assurance, CADS-dataset provides anatomically coherent supervision for reproducible dataset construction at foundation scale.

CADS-model: 
A whole-body segmentation model suite trained on the assembled CADS-dataset, illustrating how large-scale, consistently annotated data can support robust segmentation of 167 anatomical targets from head to knees. 
Evaluation on 18 public datasets and an independent hospital oncology cohort shows improved performance and generalizability across scanners, protocols, and patient populations, supported by quantitative assessment and expert qualitative review.
To support accessible use, we also release the trained models as a 3D Slicer~\cite{fedorov20123d} plugin that provides SNOMED-CT-compatible output labels~\cite{bos2006snomed}.

By combining anatomical breadth, data diversity, and modular design, CADS advances whole-body CT segmentation and provides a foundation for broader research in computational anatomy.
Beyond segmentation, CADS can support AI research tasks such as anatomical landmark detection, cross-modality registration, and anatomy-guided image synthesis.
Its anatomically comprehensive outputs may also enable applications such as longitudinal organ tracking, personalized anatomical modeling, radiation-treatment planning, and population-scale analyses of human morphology and disease.
By sharing the pre-trained models~\cite{xu2026cads}, curated large-scale data~\cite{xu2026cadsdataset}, and software~\cite{xu2026slicercads}, we aim to support robust AI development in radiology and broaden access to comprehensive anatomical analysis for clinicians and researchers.
\section*{Results} \label{sec_02_results}

Model performance is assessed on held-out splits from 18 public datasets using their original annotations, supplemented by an expert-reviewed subset of the TotalSegmentator test set to provide a more reliable benchmark in the presence of known annotation errors.
This enables assessment of structure-level accuracy across 167 targets and cross-dataset generalization. 
We further evaluate CADS in \num{2864} oncology CT scans from University Hospital Zurich, combining quantitative metrics with qualitative review by radiologists to examine its potential utility in clinical research settings.

\subsection*{CADS-dataset: A large-scale curated resource for whole-body CT analysis}
\label{sec_02_results_CADSdataset}
CADS-dataset aggregates \num{22022} CT volumes from 40 heterogeneous sources, establishing a large-scale repository for CT image analysis (Table~\ref{table:dataset_collection}).
As the data foundation of the CADS framework, it provides voxel-level annotations for 167 anatomical structures, enabling the training of a unified segmentation model.
All annotations are publicly released in their native formats for reproducibility and reuse.

Specifically, CADS-dataset consolidates data from three primary sources:
(1) large-scale public archives, including TotalSegmentator~\cite{totalsegmentator} (\num{1203} annotated volumes covering 104 anatomical structures), which provides a foundational basis for whole-body segmentation, together with the National Lung Screening Trial (NLST~\cite{nlst_data}, \num{7172} scans), CT-RATE~\cite{hamamci2024developing} (\num{3134} scans), and AbdomenCT-1K~\cite{abdomenct1k} (\num{1062} scans);
(2) specialized annotation datasets from medical imaging challenges (via Grand Challenge~\cite{grandchallenge} and MICCAI~\cite{miccaichallenge}) and focused research studies, such as HaN-Seg~\cite{hanseg} for head-and-neck organs at risk (OARs), SAROS~\cite{saros} for extended body-context structures, and several oncology studies from TCIA~\cite{tcia};
and (3) newly contributed datasets, comprising manual annotations of critical OARs on the VISCERAL~\cite{visceral} dataset, 484 newly acquired anonymized head CTs, and 586 triphasic contrast-enhanced abdominal CTs, which jointly enhance cranial coverage and multi-phase contrast representation.

The dataset’s geographic diversity spans 16 countries across multiple continents, covering data collected between 2007 and 2024 at more than 100 medical centers.
This broad representation captures a wide range of clinical practices and patient populations, spanning both healthy subjects and individuals with diverse pathological conditions, e.g., from oncologic diseases (carcinomas, lymphomas, and metastases) to COVID-19, traumatic injuries, and rare disorders.
The dataset also includes varying imaging protocols (contrast-enhanced and non-contrast acquisitions) and multiple anatomical fields of view, covering the head and neck, chest, abdomen, pelvis, spine, and whole body.

Our data curation process unifies previously fragmented resources through a pipeline that standardizes orientation, spatial resolution, and anatomical label definitions across diverse cohorts.
To support large-scale annotation development, we establish a unified anatomical label space aligned with SNOMED CT terminology, defining the scope of all 167 targets and resolving discrepancies in anatomical extent, boundary placement, and structure overlap across source datasets. 
This label space guides the integration of existing annotations and the generation of comprehensive annotations for partially labeled scans. 
Detailed label specifications and conflict-resolution criteria are provided in Supplementary Note~1.

\begin{figure*}[!htb]
    \centering
    \includegraphics[trim=0mm 0mm 0mm 0mm, clip, width=\textwidth]{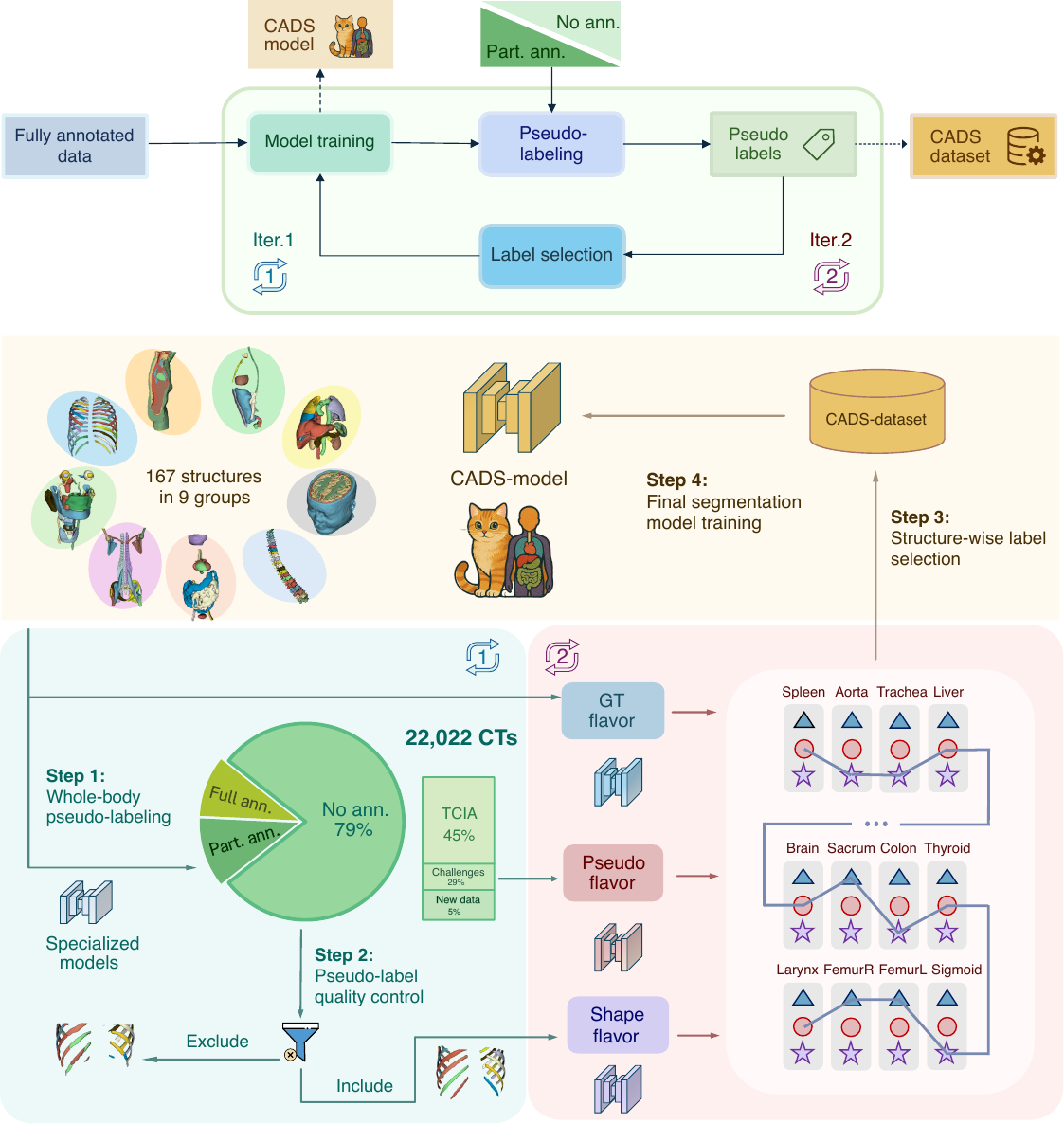}
    \caption{Development pipeline of the CADS-dataset and CADS-model.
    Top, overview of the two-iteration workflow, where models trained on fully annotated data generate pseudo-labels that are used for subsequent model training.
    Bottom, the four-step pipeline.
    Step 1: specialized models for nine model groups, together covering all 167 structures, generate whole-body pseudo-labels for the \num{22022} CT volumes.
    Step 2: a shape prior based on neural implicit functions detects implausible pseudo-labels and excludes them from training of the Shape flavor.
    Step 3: for each structure, the best-performing of three model flavors trained with different supervision (GT, Pseudo, and Shape) is selected to construct the CADS-dataset.
    Step 4: the final CADS-model is trained on the CADS-dataset with class-balanced sampling.
    Blue and pink backgrounds denote the first and second iterations.
    Full ann., fully annotated; Part. ann., partially annotated; No ann., not annotated; GT, ground truth; Iter., iteration; TCIA, The Cancer Imaging Archive.}
    \label{fig:results_overview_style_1}
\end{figure*}

Among the \num{22022} CT volumes, only \num{4695} (21\%) contain partial annotations.
To achieve full 167-structure coverage, we develop a multi-stage annotation framework (Fig.~\ref{fig:results_overview_style_1}) that converts sparse labels into comprehensive whole-body supervision.
First, target structures are grouped into nine anatomical systems, and specialized supervised models are trained using publicly available annotations.
These models then propagate pseudo-labels across unlabeled datasets, drawing on anatomical priors from diverse sources while reducing manual annotation requirements.

To further enhance label quality, we adopt a multi-expert label-fusion strategy that integrates models trained under distinct supervision regimes.
Each expert is trained on a supervision subset: original dataset-provided annotations, large-scale pseudo-labels, or pseudo-labels filtered by shape-based quality control, thereby capturing complementary anatomical priors and supervision strengths.
We design a shape-informed quality-control module based on neural implicit representations~\cite{amiranashvili2022learning,amiranashvili2024learning}, embedding each structure in a continuous shape space and estimating organ-level plausibility through signed-distance reconstruction.
Implausible pseudo-labels are automatically discarded, while anatomically consistent labels are retained.
For each anatomical target, the most reliable model variant is selected based on held-out validation, prioritizing source-mismatched out-of-distribution (OOD) validation when available and otherwise using source-matched in-distribution (ID) validation. 
Final annotations are produced through a reliability-ordered assembly that prioritizes original ground-truth annotations when available and uses predictions from the selected model variant for incompletely labeled or unlabeled scans.
This per-structure selection of model variants, guided by shape-informed quality control, yields anatomically coherent whole-body annotations while reducing systematic labeling errors without iterative manual refinement.
Further methodological details are provided in the Methods section.

Specific structures underwent anatomical refinements to improve annotation quality.
A representative example is the rib segmentation correction process, which preserves the costovertebral joints, an anatomical detail commonly missing from public datasets. 
Rather than reproducing existing dataset conventions, these refinements align annotations with anatomical definitions used in clinical practice.

\subsection*{CADS-model: A generalizable model for anatomically consistent segmentation of 167 whole-body structures} \label{sec_02_results_CADSmodel}

To evaluate the practical utility of CADS-dataset, we train CADS-model as a standardized, region-specific nnU-Net~\cite{isensee2021nnu} implementation covering 167 anatomical targets from head to knees (Fig.~\ref{fig:results_whole_body_parts_visualization}). 
The contribution is not a new network architecture but a released model demonstrating how large-scale, consistently curated annotations improve whole-body segmentation within an established framework.
For computational efficiency and memory management, the 167 targets are grouped into anatomical regions, each handled by a dedicated U-Net~\cite{ronneberger2015u}.
All components use the same training split of CADS-dataset under consistent preprocessing, label definitions, and optimization settings.
This differs from approaches that combine models trained on separate datasets with heterogeneous label conventions~\cite{totalsegmentator}. 
To support reproducible evaluation and downstream use, we release the model weights together with a user-friendly 3D Slicer plugin.

\subsection*{Anatomical precision across the body: Structure-level performance analysis} \label{sec_02_results_per_structure}
\begin{figure*}[!htbp]
    \centering
    \includegraphics[trim=0mm 0mm 0mm 0mm, clip, width=\textwidth]{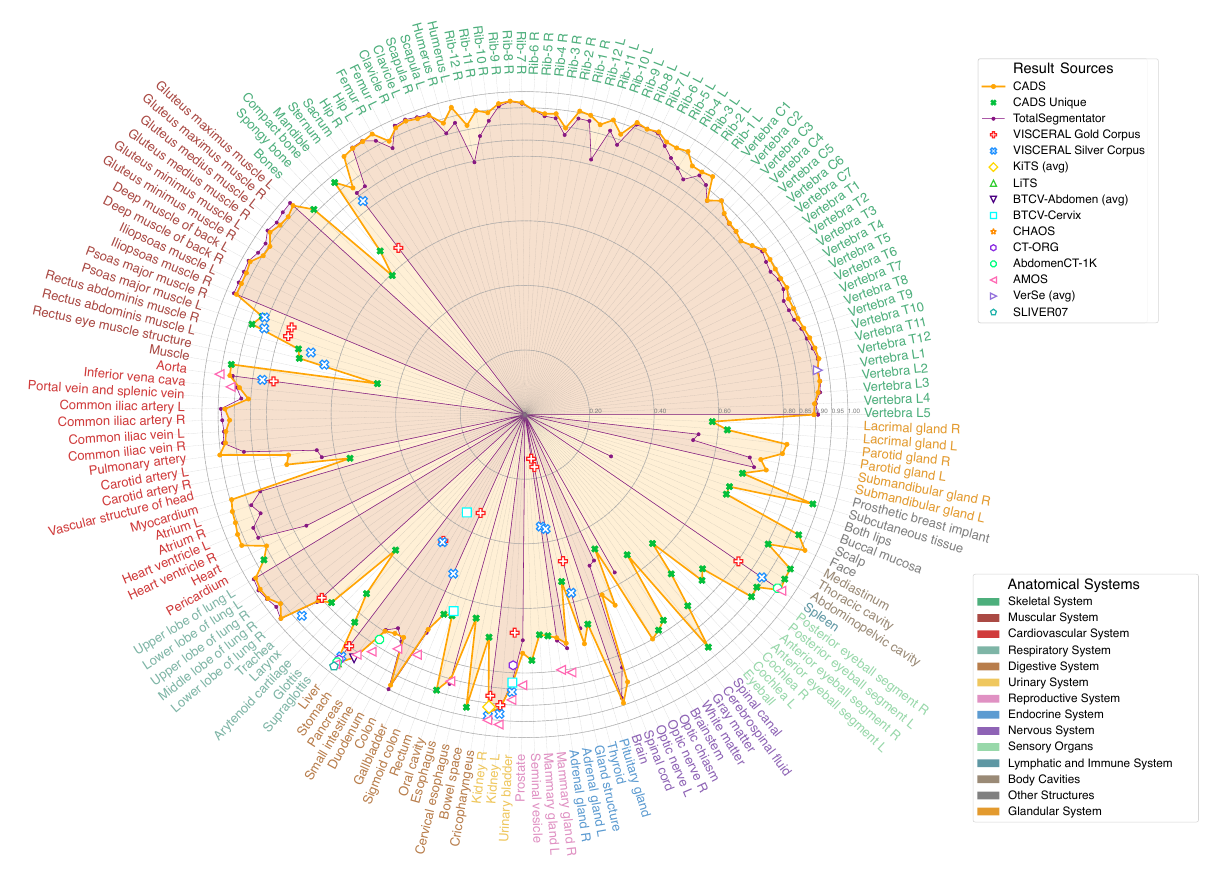}
    \caption{Structure-level segmentation performance.
    Mean Dice scores of the CADS-model (orange) and TotalSegmentator (purple) across 167 structures, evaluated on the test sets of 18 public datasets. 
    Structure labels are colored by anatomical system, and the radial axis ranges from 0 at the center to 1 at the periphery.
    Outlined symbols indicate published results from challenge-winning methods on the corresponding datasets; (avg) indicates scores averaged across multiple structures.
    Green crosses mark the 49 structures available only in the CADS-model; their TotalSegmentator scores are shown as 0.}
    \label{fig:results_radar_dice}
\end{figure*}

To evaluate CADS-model at the structure level, we analyze individual anatomical structures using 18 public datasets that provide ground-truth annotations for specific targets (data partitions in Supplementary Table~6), so that every structure is tested against multiple independent sources spanning diverse imaging protocols and patient populations.
Fig.~\ref{fig:results_radar_dice} summarizes structure-wise Dice scores for all 167 targets relative to existing approaches.

To isolate the impact of training data from architectural differences, we select TotalSegmentator~\cite{totalsegmentator} as the primary baseline, because it uses the same nnU-Net framework~\cite{isensee2021nnu} and enables a direct comparison of dataset contributions.
This is particularly relevant because several whole-body segmentation methods, including MONAI Label~\cite{diaz2024monai} and other recent approaches~\cite{plotka2024swin,ji2023continual}, have been intensively developed using the TotalSegmentator dataset.
We evaluate TotalSegmentator v2.4.0 using its strongest available inference configuration, including robust region-of-interest refinement and specialized sub-task models where available. 
Full details are provided in Supplementary Note~5.
For additional context, we include winning methods from organ-specific segmentation challenges using their originally reported metrics, with outlined symbols in Fig.~\ref{fig:results_radar_dice}.
For each structure, performance is calculated only on test samples with consistent anatomical definitions and ground-truth annotations.

\begin{figure*}[!htb]
    \centering
    \includegraphics[trim=0mm 0mm 0mm 0mm, clip, width=\textwidth]{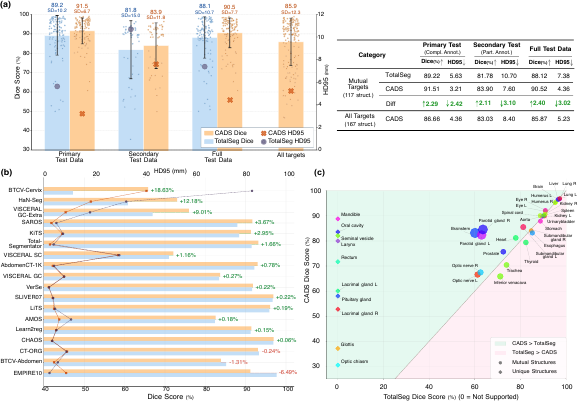}
    \caption{Benchmark and clinical evaluation of CADS-model performance.
    (a) CADS-model and TotalSegmentator on primary test data with complete annotations, secondary test data with partial annotations, and the full test data. CADS-model performance is also shown across all 166 targets, excluding the face because of de-identification.
    Bars show macro-averaged Dice (dots, per-structure means; error bars, standard deviation across structures; n = 117, 50, 117, and 166 structures, respectively), and markers show macro-averaged HD95 (right axis). The table reports the corresponding values; arrows indicate whether higher ($\uparrow$) or lower ($\downarrow$) values are better.
    (b) Dataset-level mean Dice (bars, bottom axis) and HD95 (lines, top axis) across the 18 public test datasets. Numbers indicate the Dice difference between CADS-model and TotalSegmentator in percentage points (green, CADS-model higher; red, CADS-model lower).
    (c) Mean Dice of CADS-model versus TotalSegmentator for 38 structures with clinical reference annotations in the University Hospital Zurich cohort (\num{2864} CT scans). Circles indicate the 28 structures supported by both models, with size proportional to the absolute Dice difference; diamonds denote the 10 structures supported only by CADS-model. Shading refers to which model has the higher Dice score; colors distinguish individual structures.
    HD95, 95th-percentile Hausdorff distance; Compl. Annot., complete annotations; Part. Annot., partial annotations; struct., structures.}
\label{fig:results_quantitative_comparisons}
\end{figure*}

Results demonstrate the benefits of training on the larger, more diverse CADS-dataset rather than on the TotalSegmentator dataset.
For the 117 mutual targets (excluding the face class), CADS-model achieves a macro-averaged mean Dice score of \num{90.52}\% (\num{95}\% CI: \num{88.12}\%--\num{92.41}\%; median: \num{92.33}\%), compared with \num{88.12}\% (\num{95}\% CI: \num{85.18}\%--\num{90.48}\%; median: \num{90.13}\%) for the baseline model with a similar nnU-Net architecture trained on the limited dataset (Fig.~\ref{fig:results_quantitative_comparisons}-a).
Training on the CADS-dataset achieves higher mean Dice scores in 71 structures, including 44 with $p < \num{0.05}$ in the structure-wise paired comparisons.
Challenge leaderboard results show performance fluctuations across anatomical structures, reflecting methods trained on single-source or otherwise limited datasets and optimized for different targets in earlier years. 
By contrast, training on the large-scale, diverse CADS-dataset yields more consistent performance across the anatomical spectrum.

The structure-wise results show that improvements are not uniform across organs. 
Gains are most apparent for targets with limited prior coverage, complex boundaries, or greater sensitivity to annotation conventions, whereas several large, high-contrast organs show smaller differences from existing state-of-the-art models. 
This pattern is consistent with the broader aim of CADS to improve whole-body anatomical coverage beyond well-established organ targets.

Beyond the mutual targets, the CADS-model segments 49 additional structures, shown by green cross markers in Fig.~\ref{fig:results_radar_dice}. 
These include critical sensory organs, body cavities, and reproductive structures.
Across all 167 target structures, the CADS-model achieves an overall Dice score of \num{85.87}\%, which is slightly lower than in the mutual-target comparison because several added structures pose specific segmentation challenges:
(1) limited training data for certain structures, as illustrated by the buccal mucosa, which appears in fewer than \num{1000} scans with only 30 ground-truth annotations, compared with well-represented structures present in more than \num{10000} scans with several hundred ground-truth annotations (Table~\ref{table:cads-dataset_statistics});
(2) very small organs (volumes below \SI{0.5}{\milli\litre}), such as the arytenoid cartilage, where standard CT resolution and partial-volume effects limit accuracy;
and (3) anatomically complex structures, particularly small glands.
The three categories identify targets for future refinement. 
Complete performance metrics beyond Dice scores are provided in Supplementary Note~5.

Taken together, these results show consistent performance across both shared and CADS-only targets.
Because CADS differs from the baseline in dataset scale, anatomical coverage, source diversity, and label curation, these improvements should not be attributed to training-set size alone.
Additional analyses of dataset quantity, diversity, and annotation quality are provided in Supplementary Note~4.

\subsection*{Cross-dataset generalization: Dataset-level performance analysis} \label{sec_02_results_per_dataset}
We evaluate CADS-model across individual data sources using heterogeneous test sets (Fig.~\ref{fig:results_quantitative_comparisons}-b).
We report the Dice coefficient for volume overlap and the 95th-percentile Hausdorff distance (HD95) for boundary precision; further measures are given in Supplementary Note~5.

Across the 18 test datasets, CADS-model shows improved performance compared with baseline TotalSegmentator in most cases, with higher Dice scores in 15 datasets and lower HD95 values in 16 datasets.
On average across the 18 datasets, CADS-model achieves a Dice score \num{2.40} percentage points higher and an HD95 \SI{3.96}{\milli\meter} lower.
Notable improvements in boundary precision are observed in anatomically complex regions, with substantial HD95 reductions in BTCV-Cervix (\SI{42.21}{\milli\meter}) and HaN-Seg (\SI{14.18}{\milli\meter}).

For a more nuanced evaluation, we stratify our analysis into two cohorts:
(1) a primary cohort of held-out splits from sources represented during development, with complete ground-truth annotations for all structures, providing the most favorable benchmark for accuracy assessment; and
(2) a secondary cohort with incomplete or selective labels, reflecting real-world scenarios where annotations often cover only specific structures of interest. 
Training on CADS-dataset yields consistent improvements across both cohorts, with Dice score increases (primary: \SI{+2.29}{\percent}, secondary: \SI{+2.11}{\percent}, full dataset: \SI{+2.40}{\percent}) and HD95 metric improvements (reductions of \SI{2.42}{\milli\meter}, \SI{3.10}{\milli\meter}, and \SI{3.02}{\milli\meter}, respectively).
Gains of similar size under both annotation regimes suggest adaptability to varying levels of ground-truth completeness, which is particularly relevant for clinical research settings.

Evaluation on the TotalSegmentator dataset, which provides reference annotations for 104 structures, reveals annotation inaccuracies across multiple structures, most prominently in ribs and vertebrae (Fig.~\ref{fig:results_visualization}-a), with additional examples documented in Supplementary Note~5.
These annotation inaccuracies may affect both the evaluation of segmentation performance and models trained on this dataset, underscoring the importance of corrected annotations for future benchmarking and model development.
To account for these annotation quality concerns, we create an expert-verified subset for more reliable benchmarking.
On this curated test set, our model trained on the CADS-dataset achieves a Dice score of \num{93.15}\% and improves the delineation of structures such as the costovertebral joints, where the original labels contain errors (Fig.~\ref{fig:results_visualization}-a). 
\subsection*{Real-world evaluation of the CADS-model in a hospital oncology cohort} \label{sec_02_results_clinical_applicability}
While automated segmentation has shown promise on curated public datasets, its clinical value depends on performance in pathological cases encountered in routine hospital practice.
To assess this setting, we evaluate our model on \num{2864} subjects from the Radiation Oncology Department at University Hospital Zurich, representing an independent cohort that is not involved in model development.
This cohort spans 38 anatomical structures and diverse pathological conditions, including kidney and liver tumors, lung cancer and other malignancies, reflecting challenges that AI systems may encounter in clinical practice.
Detailed characteristics of this cohort are provided in the Methods section.

In this real-world evaluation, training on the CADS-dataset improves performance across most structures (Fig.~\ref{fig:results_quantitative_comparisons}-c; bubble size indicates Dice score differences).
The largest gains are observed in structures relevant to radiation therapy planning, including the brainstem (\num{23.08} percentage points) and the parotid glands (left: \num{19.40}; right: \num{21.02}).
Accurate delineation of these structures is important for radiation dose planning and normal-tissue sparing.
These gains may partly reflect the broader representation of radiation-oncology-relevant structures in CADS.

To evaluate clinical relevance beyond quantitative metrics, we conduct a systematic expert review led by a radiation oncologist with 5.5 years of experience.
For each structure, the expert assesses three cases around the median Dice score: one at the median and two at adjacent performance levels, so that the review reflects typical rather than best-case performance.
Clinical acceptability is assessed through slice‑by‑slice review of the 3D segmentation masks, considering geometric accuracy, anatomical completeness and consistency with consensus organ-at-risk delineation practices~\cite{mir2020organ}.
Expert assessment indicates that CADS-model segmentations are acceptable for anatomical structure delineation across most structures in radiation-therapy planning scenarios (Fig.~\ref{fig:results_visualization}-b, lower part). 
The model demonstrates particular strength in traditionally challenging regions such as the parotid deep lobe, mandible, and complex lung areas (Fig.~\ref{fig:results_visualization}-b, upper part).
Nevertheless, the expert review also notes areas for improvement, including label-definition differences for the brain and trachea and imprecise delineation of the optic chiasm and inferior vena cava (Fig.~\ref{fig:results_visualization}-c).
Together with the quantitative results, this qualitative assessment suggests potential relevance for radiation-oncology contouring workflows, but does not constitute a formal demonstration of clinical utility, which would require blinded multi-reader studies comparing CADS-generated contours with manual clinical annotations.

Validated through both quantitative metrics and expert visual assessment, the results support the suitability of our data-centric approach for RT-planning research.
The performance of both AI models indicates that machine learning-based medical image segmentation continues to improve in clinically relevant tasks, with comprehensive, high-quality datasets playing an increasingly important role alongside architectural advances.

\begin{figure*}[!htb]
    \centering
    \includegraphics[trim=0mm 0mm 0mm 0mm, clip, width=\textwidth]{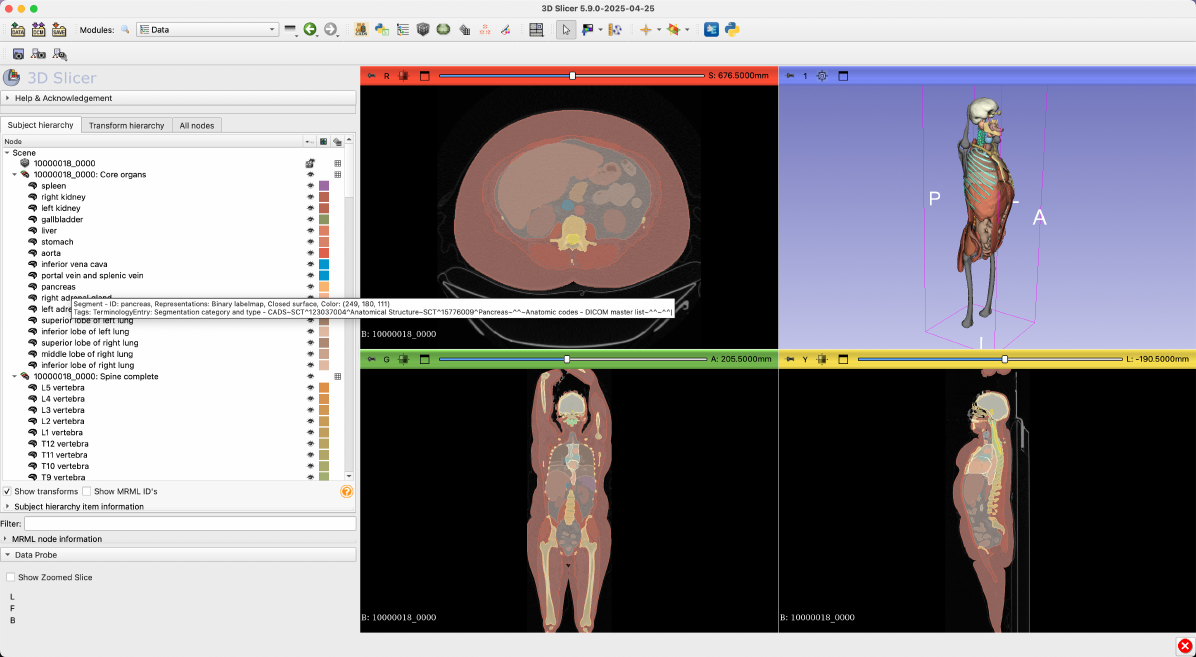}
    \caption{CADS in 3D Slicer.
    The CADS extension in 3D Slicer~\cite{fedorov20123d}, showing whole-body segmentation of a de-identified CT scan from the VISCERAL Gold Corpus~\cite{visceral} in axial, coronal, and sagittal views, together with a three-dimensional rendering.
    Segments are grouped by model group in the data hierarchy (left) and linked to their corresponding SNOMED CT codes.
    Colors distinguish individual structures.}
    \label{fig:3dslicer}
\end{figure*}

Several platforms and tools, including 3D Slicer~\cite{fedorov20123d}, OHIF~\cite{ziegler2020open}, and MONAI Label~\cite{diaz2024monai}, provide user-friendly interfaces that enable clinicians to use AI algorithms with minimal technical expertise. 
To broaden accessibility, we release a 3D Slicer plugin implementing the CADS-model (Fig.~\ref{fig:3dslicer}), providing whole-body segmentation within an interface already familiar to many clinicians and supporting research use of AI-assisted segmentation workflows.

\section*{Discussion} \label{sec_03_discussion}
AI-based whole-body CT analysis is reaching a point where technical advances increasingly meet clinical needs.
However, progress in whole-body segmentation remains hindered by fragmented data, limited anatomical coverage, and insufficient clinical validation.
A central barrier is the lack of comprehensively annotated datasets, which impedes the development of robust segmentation models with potential clinical utility.

The CADS framework addresses this gap through a data-centric approach. 
Rather than pursuing incremental optimizations in model architecture, we prioritize the systematic integration, standardization, and labeling of heterogeneous data sources.
The resulting CADS-dataset comprises \num{22022} CT volumes annotated across 167 structures (Fig.~\ref{fig:results_whole_body_parts_visualization}), spanning multiple continents, protocols, and pathologies.
This scale and diversity provide a foundation for developing models with robust cross-institutional generalization.
Using the established nnU-Net architecture, we show that training on CADS improves performance over models trained with more limited anatomical coverage and data diversity.
Our open-source 3D Slicer plugin (Fig.~\ref{fig:3dslicer}) further makes these capabilities accessible to clinicians in familiar environments, while SNOMED CT terminology integration supports their incorporation into clinical research workflows.

Throughout this study, clinical feedback informs dataset curation, benchmark assessment, and qualitative model evaluation. 
During CADS-dataset construction, annotation issues identified during model development, particularly in anatomically complex structures such as ribs and vertebrae, prompt large-scale label refinements and quality-control procedures.
These findings show that label-quality issues in widely used public datasets can remain unnoticed and may affect both model training and benchmark evaluation.
For benchmark assessment, the held-out TotalSegmentator test subset is reviewed at the case-structure level; unreliable reference labels are excluded from metric computation, and a corrected mini-benchmark is created by manually revising a subset of the excluded labels and re-evaluating both CADS and TotalSegmentator (Supplementary Note~5).
Finally, in the independent University Hospital Zurich oncology cohort, expert clinical review of representative CADS predictions highlights both strengths and directions for future improvement.

This data-centric strategy also requires an annotation pipeline that can scale while maintaining sufficient consistency across heterogeneous sources.
Our large-scale automated annotation approach aims to balance the inherent trade-off between precision and scalability. 
Pseudo-label errors arise from boundary ambiguity, protocol heterogeneity, and inter-annotator inconsistency, and have both random and systematic components.
The main advantage of our approach lies in reducing the random, variance-driven component, which is common in pseudo-labeling and label-propagation settings with limited labeled data~\cite{shen2023balancing,feofanov2024multi}.
Averaging predictions across models trained on different supervision subsets reduces the variance of the resulting labels, with the achievable reduction limited by the correlation between those models.
Shape-based outlier detection complements this by withholding implausible pseudo-labels from the shape-based model variant.
The overall improvements across 18 test datasets (Fig.~\ref{fig:results_quantitative_comparisons}) are compatible with this reduction in label variance, although they also reflect the greater scale, coverage, and diversity of the training data.
To address the systematic component, CADS applies targeted anatomical refinements, such as costovertebral joint preservation, to correct omissions common in existing datasets.

CADS should be viewed not as competing with recent algorithmic innovations, but as a complementary foundation for extending their impact.
Many state-of-the-art models~\cite{pai2025vision,liu2023clip,lei2025medlsam,huang2023stu} have been developed or evaluated on datasets with the limitations identified in our analysis, potentially constraining their performance.
By providing annotations at scale within a clearly defined and unified anatomical label space, CADS offers a practical route to further improve these models through training on more comprehensive data.

Despite these strengths, CADS has limitations that point to directions for future research.
First, our shape-based quality assessment withholds low-quality pseudo-labels from training the shape-based model variant.
Although this strategy supports annotation consistency at the scale of \num{22022} CT volumes, it may also filter out labels for clinically important pathological or atypical anatomy.
Manual correction would better preserve such labels but is not feasible at this scale.
Future work should therefore explore hybrid quality-control strategies that combine automated filtering with targeted manual review to retain difficult but informative training instances.
Second, our real-world evaluation includes qualitative review by a radiation oncologist, but this assessment is performed by a single non-blinded expert and is designed to identify strengths and areas for improvement rather than to establish clinical utility. 
More rigorous clinical evaluation should include multiple reviewers, source-blinded assessment protocols, and clinically defined endpoints.

Third, although CADS covers anatomy from head to knees, its label space could be expanded to additional regions, such as the hands and feet, and to more fine-grained anatomical structures. 
This is particularly relevant for the central nervous, musculoskeletal, and vascular systems, where detailed structures are often absent from current datasets because of annotation complexity.
Several existing labels also require refinement:
the brain mask includes all central nervous system components within the skull, which may be too inclusive for certain applications;
the trachea mask covers the airway lumen but not the surrounding cartilage;
the optic chiasm tends to be under-segmented;
and the rectum is shorter than its anatomical extent.
In addition, facial anonymization limits applications that need detailed facial analysis; these would require complementary datasets or hybrid approaches.
Finally, future methodological work could explore unified architectures that segment all structures simultaneously while preserving the efficiency and coverage of the current modular design. 

By publicly sharing the CADS framework, including the CADS-dataset~\cite{xu2026cadsdataset}, CADS-model~\cite{xu2026cads}, and a ready-to-use tool~\cite{xu2026slicercads}, we aim to facilitate further research in medical image analysis.
The anatomical hierarchies encoded in CADS may support context-aware segmentation strategies that exploit spatial relationships between organs, while the dataset itself enables tasks beyond segmentation, including anatomical landmark detection, cross-modality registration, and anatomy-guided image reconstruction.
Potential applications span radiation oncology, where rapid organ delineation can streamline treatment planning; diagnostic radiology, where automated organ volume quantification may help detect subtle anatomical changes; and population-level research across diverse conditions and cohorts.
Together, these resources provide a basis for studying how large-scale segmentation methods can support clinical research and AI-integrated image analysis workflows.

\clearpage 
\section*{Methods} \label{sec_04_methods}

\subsection*{Ethics statement}
\label{sec_04_methods_ethics_statement}
This research complies with all relevant ethical regulations.
The Clinical Research Ethics Committee of Istanbul Medipol University approved the collection, scientific use, and publication of 484 head CT scans and their associated annotations and derived models (Approval No. E-10840098-772.02-6841; October 27, 2023).
The Ethics Committee of the Technical University of Munich approved the scientific use and publication of 586 anonymized triphasic contrast-enhanced abdominal CT scans (Approval No. 27/19 S-SR).
All remaining data were obtained from publicly available datasets under their respective licenses, for which no additional ethical approval was required.

\subsection*{Dataset collection and preprocessing}
\label{sec_04_methods_collection_preprocessing}
We aggregate CT scans from medical imaging challenges hosted on Grand Challenge~\cite{grandchallenge} or organized at MICCAI~\cite{miccaichallenge}, public repositories such as The Cancer Imaging Archive (TCIA)~\cite{tcia}, and two newly collected hospital cohorts.
Together, these scans cover major anatomical regions from head to knees, including the head and neck, thorax, abdomen, pelvis, and thighs.
To maximize scale and diversity, we do not restrict inclusion to annotated scans; approximately 79\% of the collection therefore has no annotations.
To strengthen the annotated portion, we incorporate existing large-scale annotated datasets such as TotalSegmentator~\cite{totalsegmentator} and SAROS~\cite{saros}.

The final CADS-dataset contains \num{22022} CT scans, making it one of the largest whole-body CT collections for segmentation to date (Table~\ref{table:dataset_collection}). 
Scans are partitioned into non-overlapping development and test sets (Supplementary Table~6).
All model development, including hyperparameter tuning and architecture selection, is performed on validation cases separated from the training cases within the development set, so that the test sets remain fully independent.

To ensure consistency across images acquired from different sources, we apply a standardized preprocessing pipeline.
All scans are first reoriented to the RAS (Right-Anterior-Superior) convention and resampled to \SI{1.5}{mm} isotropic resolution, balancing anatomical detail with the memory and runtime requirements of deep learning models. 
We then simplify the affine transformation matrices by removing rotation and translation components while preserving scaling and shear.
This enables consistent downstream analyses such as pseudo-label quality control, which relies on shape analysis and direct comparison of anatomical structures across scans.
For the newly collected head CT scans, facial regions are blurred with a Gaussian filter ($\sigma = 5$ voxels at \SI{1.5}{mm} spacing) to reduce identifiability while preserving nearby anatomical context.

\subsection*{CADS-dataset annotation pipeline}
\label{sec_04_methods_annotation_pipeline_development}

\subsubsection*{From partial annotations to whole-body pseudo-labels}
\label{sec_04_methods_stage_1_wholebody_pseudolabling}
Because the collected dataset is only partially annotated, with annotations available for 21\% of scans and often limited to one or two structures, we use pseudo-labeling to extend supervision to all 167 target structures across the full \num{22022}-scan collection~\cite{chapelle2009semi}.
We organize the target structures into nine anatomical groups (Step 1 in Fig.~\ref{fig:results_overview_style_1}; group definitions in Fig.~\ref{fig:results_whole_body_parts_visualization}) and compile initial training data from five sources:
(i) TotalSegmentator~\cite{totalsegmentator}, which provides labels for 104 structures, including major organs, vessels, and bones, and forms five of the nine groups;
(ii) in-house annotations of 10 radiotherapy organs at risk, including the spinal canal, larynx, and heart, contoured on VISCERAL Gold Corpus images~\cite{visceral} by three medical research assistants and a radiation oncologist (J.C.P.);
(iii) HaN-Seg~\cite{hanseg}, which provides head-and-neck structures such as the cochlea, optic nerves, and carotid arteries;
(iv) brain-structure labels generated by aligning HaN-Seg CT scans to the SimNIBS MRI atlas~\cite{simnbis} using ANTs non-rigid registration~\cite{avants2009advanced};
and (v) SAROS~\cite{saros}, which provides broad anatomical context structures, including subcutaneous tissue and major body cavities.
Because SAROS provides sparse 2D annotations on selected slices, we interpolate between annotated slices to obtain dense 3D labels for volumetric training.

To handle the anatomical heterogeneity across groups, we train a specialized nnU-Net model~\cite{isensee2021nnu} for each group on these initial training sets and use the resulting models to generate pseudo-labels for all target structures across the collection.

\subsubsection*{Shape-informed label quality control}
\label{sec_04_methods_stage_2_quality_control}
The diversity of images in our large-scale CT collection is a key property of our dataset and is a major contributor to robustness of the emerging segmentation models.
A dataset of this scale also inevitably contains outlier images -- volumes with wrong spacings and orientation, unusually high noise levels, low contrast, etc.
Such outliers cause catastrophically bad segmentations during pseudo-labeling.
While the overall image diversity is important for downstream model robustness, erroneous pseudo-labels can degrade the final model's performance~\cite{zhang2020weakly,kwon2022semi}.
We balance this trade-off by ranking pseudo-labels in terms of their quality and removing a certain percentage of worst pseudo-labels from the downstream training data.

We formulate ranking of pseudo-labels as likelihood estimation in shape space.
Our approach has a few unique properties.
First, it is automated, allowing ranking of large-scale datasets.
Second, training of our likelihood estimator requires only ground-truth segmentations and does not rely on degraded shapes at training time.
This makes our method's ranking performance agnostic to any specific failure modes of the pseudo-labeling network, essentially performing unbiased, zero-shot (or unsupervised) outlier detection.
Lastly, we focus directly on pseudo-labels instead of images for ranking.
Arguably, this leads to a simpler problem outlier detection in shape space, compared to image space.
Furthermore, it allows us to take the generalization ability of the pseudo-labeler network into account and only exclude images where it fails.

To perform likelihood estimation, we train an auto-decoder on ground-truth multi-class segmentation labels from the fully annotated datasets.
The auto-decoder is based on implicit neural representations~\cite{amiranashvili2022learning,amiranashvili2024learning} and allows us to effectively model heterogeneous high-resolution volumes with varying fields of view.
During training, the decoder weights and a latent code for each training shape are optimized jointly; at inference, the decoder is kept fixed and a new latent code is optimized so that the decoded segmentation best matches the given pseudo-label~\cite{amiranashvili2022learning,amiranashvili2024learning}.
Since the shape prior has been trained on clean ground-truth segmentation masks, it is less able to reconstruct erroneous segmentation masks.
Therefore, shape likelihood can be connected to the distance between the observed pseudo-label and its reconstruction, with large distance indicating an outlier.
Note that this method does not rely on any ground-truth at inference time.

Different choices of the metric can be made to measure the distance between a pseudo-label and its reconstruction.
We choose Hausdorff distance due to its sensitivity to outliers, in contrast to Dice score.
Since a few far-off voxels are unlikely to hinder subsequent segmentation model training, we choose a more robust 90-percentile Hausdorff distance.
This approach allows us to automatically estimate quality scores for pseudo-labels without relying on their corresponding ground-truth, with a large Hausdorff distance indicating a low-quality pseudo-label.

\subsubsection*{Complementary model flavors for structure-wise annotation}
\label{sec_04_methods_3_flavors}
The specialized models are initially trained on small, fully annotated datasets for specific anatomical structures.
When applied to the full \num{22022}-scan collection, these models may be affected by distribution shifts between source and target data~\cite{guan2021domain,yu2023source} and by the limited anatomical variation captured in their initial training sets~\cite{liu2021s,maynord2023semi}, both of which can introduce errors into the pseudo-labels.
To improve label quality across the CADS-dataset, we use a curation-time, structure-wise multi-model annotation approach (Step 3 in Fig.~\ref{fig:results_overview_style_1}) with three components:
preparing distinct labeled subsets, training complementary models on these subsets, and assembling final labels through structure-wise model selection.
We refer to the three complementary models as flavors.

The GT-flavor model is trained directly on the original ground-truth annotations.
Although this subset is smaller than the pseudo-labeled training set, it provides the highest-fidelity supervision available without further manual intervention.
This flavor is therefore useful when preserving precise anatomical definitions and annotation conventions is more important than maximizing training-set diversity, as for brain tissue classes and cavity or tissue-composition labels.
Its main limitation is the restricted anatomical, pathological, protocol, and field-of-view variation represented by the available paired image-label samples.

The Pseudo-flavor model is trained on the full training split using the initial pseudo-labels.
The large volume of pseudo-labeled data helps the model learn the underlying data distribution and reduces the risk of overfitting compared with training on smaller datasets, consistent with the cluster assumption of semi-supervised learning~\cite{chapelle2009semi}.
Although this flavor benefits from scale and diversity, it can remain sensitive to pseudo-label noise, including boundary errors, leakage, and anatomically implausible masks.

The Shape-flavor model is trained on pseudo-labels retained after shape-informed quality control.
The shape prior described above assigns a plausibility score to each eligible structure-specific pseudo-label, and the 10\% of images with the lowest overall scores are withheld from the Shape-flavor training set; these images remain in the Pseudo-flavor training set and in the final CADS-dataset.
Compared with the Pseudo-flavor, this flavor reduces the influence of implausible masks, including severe fragmentation, leakage, discontinuity, and partial-field-of-view artifacts.
It is particularly useful for structures sensitive to outlier masks or topology errors, such as vessels and other tubular structures, the bowel, and pelvic soft tissues.
The 10\% threshold is chosen to improve pseudo-label reliability while retaining sufficient training samples for robust learning; an analysis of its sensitivity and of the filter's behavior on atypical anatomy is provided in Supplementary Note~2.

\subsubsection*{Structure-wise flavor selection}
\label{sec_04_methods_flavor_ranking}
After training the three flavors, we select the most reliable one for each target structure using held-out validation data.
Performance is assessed on both in-distribution (ID) and out-of-distribution (OOD) validation sets.
ID validation cases are held out from the same five core data sources used for initial model training, whereas OOD validation cases are drawn from other labeled datasets in our collection.
For example, liver predictions from models trained on TotalSegmentator are additionally evaluated against liver annotations from LiTS, VISCERAL, and BTCV-Abdomen.

For each structure, flavors are ranked primarily by the Dice similarity coefficient (Dice), computed on OOD validation cases where available, because these more directly assess cross-dataset generalization, and on ID cases otherwise.
Flavors are compared by statistical testing at $p < 0.05$, and those without significant differences are assigned tied ranks, which are resolved using the 95th-percentile Hausdorff distance (HD95).
The highest-ranked flavor is selected as the model-generated label source for that structure in the CADS-dataset.
The full ranking procedure, including the statistical tests used, the selected flavor for each structure, and summaries of the validation regimes, is provided in Supplementary Note~2.

\subsubsection*{Assembly of the final label set}
\label{sec_04_methods_assemble}
The final CADS label set is assembled from the pseudo-labels of the selected flavor for each anatomical structure, with original ground-truth annotations taking priority wherever they are available.
Flavor selection therefore determines the final label only for structures that lack original annotations in a given scan.

To resolve overlaps between structures within each anatomical group, whose labels may come from different flavors, we use a progressive merging strategy based on label reliability.
Within each group, structures are merged in ascending order of mean Dice, so that more reliable labels overwrite less reliable ones.
Structures with equal mean Dice are ordered by the proportion of ground-truth annotations in their training data, with higher proportions merged later.
Original ground-truth annotations are merged last and overwrite any existing pseudo-labels for the same structures.
Summary statistics of the resulting label set are provided in Table~\ref{table:cads-dataset_statistics}.
Supporting analyses, including input-adaptive routing as an extension of curation-time flavor selection and internal ablations of each development stage, are provided in Supplementary Notes~2 and 4, respectively.

\subsubsection*{Specification of the CADS label space}
\label{sec_04_methods_labelspace_specification}
The 167 segmentation targets in the CADS-dataset are not a simple union of labels from different sources.
Instead, they form a curated label space in which mutually exclusive, hierarchical, and overlapping labels are explicitly distinguished.
Starting from the available ground-truth annotations, we review each source label with respect to its anatomical scope, field-of-view completeness, and relation to other released targets.
Through this review, we merge equivalent labels, resolve source-specific ambiguities, and retain semantically distinct coarse and fine-grained structures as related targets when they support different downstream uses.

For each released target, we specify its SNOMED CT identifier, related anatomical metadata, and the annotation guideline or source convention used to define its contour (Supplementary Table~1).
Because the label space is hierarchical, coarse and fine-grained targets can coexist in different model groups and may overlap.
We therefore document containment and overlap relations across groups and verify them with a post hoc spatial-consistency audit.
Details of the label-space specification, cross-source conflict types, integration rules, the spatial-consistency audit, and the body-region-based anatomical hierarchy are provided in Supplementary Note~1.

\subsection*{CADS-model training}
\label{sec_04_methods_cads_model_architecture}

\subsubsection*{Class-balanced sampling and region-specific training}
\label{sec_04_methods_cads_model_training}

Following the construction of the CADS-dataset, we train the CADS-model to exploit its scale and anatomical diversity and to improve generalization relative to models trained on smaller, single-source cohorts.
We use the residual-encoder U-Net configuration (ResEnc-L) of nnU-Net~\cite{isensee2024nnu} (v2.5.1), given its strong performance in medical image segmentation and its suitability for large-scale training.
A challenge in this setting is the uneven representation of anatomical structures across the training volumes, which can bias optimization toward frequently observed classes~\cite{li2020analyzing,zhong2023understanding,huang2016learning}.
In the CADS-dataset, for example, the upper cervical vertebrae (C1--C5) each appear in only 905 to \num{3255} of the \num{22022} volumes, whereas the liver, spleen, and aorta each appear in more than \num{21000} (Table~\ref{table:cads-dataset_statistics}).

To reduce this bias and promote more balanced performance across the 167 target structures, we modify the standard nnU-Net sampling strategy.
Although nnU-Net combines uniform image sampling with foreground-enriched patch extraction, it does not explicitly balance sampling among foreground classes.
We therefore assign sampling weights based on the inverse probability of class occurrence, increasing the chance of selecting images that contain under-represented structures; for instance, volumes containing C1--C5 receive higher sampling weights than those containing only more frequently observed vertebrae.

We further define region-specific training subsets for the CADS-model groups, including the brain and head-and-neck organ-at-risk groups, to improve training efficiency while preserving target-region coverage.
The subset-selection criteria and training image counts for each group are provided in Supplementary Note~2, and structure-specific post-processing, including head-region cropping and rib label refinement, is described in Supplementary Note~3.

\subsubsection*{Complementary CADS-model releases}
In addition to the reference CADS-model evaluated in this study, we release complementary variants for different redistribution terms and application requirements.
These include licensing-stratified CADS-models, trained on subsets of the CADS-dataset restricted to progressively more permissive licenses, and CADS-region, a body-region-driven model suite that reorganizes the CADS-dataset into region-specific tasks.
The rationale, training composition, and evaluation of these releases are described in Supplementary Note~3.

\subsubsection*{Architecture benchmark on the CADS-dataset}

To assess the sensitivity of CADS-model performance to network architecture, we benchmark representative CADS target groups under a controlled setting with the same training data, preprocessing, nnU-Net plans, and held-out evaluation split.
The benchmark compares the ResEnc-L U-Net used for the reference CADS-model with convolutional (MedNeXt-M~\cite{roy2023mednext} and SegResNet~\cite{myronenko20183d}), transformer-based (SwinUNETR~\cite{hatamizadeh2021swin}), and state-space (U-Mamba Bot~\cite{ma2024u}) architectures.
Its design, target groups, and results are described in Supplementary Note~3.

\subsection*{Evaluation}
\label{sec_04_methods_evaluation}

\subsubsection*{Evaluation metrics and false-negative penalization}
\label{sec_04_methods_evaluation_metrics}

Segmentation performance is evaluated using Dice as the primary overlap metric and HD95 as the primary boundary metric. 
Additional metrics, including the Hausdorff distance (HD), normalized surface Dice (NSD), true-positive rate (TPR), and error volume, are used in extended analyses.
Cases with a non-empty ground-truth segmentation but an empty prediction are treated as false negatives: Dice and TPR are assigned a value of zero, and distance-based metrics are assigned a maximum-distance penalty. 
Structures with a ground-truth volume below 10\% of the corresponding median organ volume are excluded from evaluation, as these generally represent organs located predominantly outside the imaging field of view. 
Full metric definitions and rules for handling empty segmentations are provided in Supplementary Note~5, and statistical analyses are described under Statistics and reproducibility below.

\subsubsection*{TotalSegmentator reference-label review and corrected mini-benchmark}
\label{sec_04_methods_review_ts_minibenchmark}
TotalSegmentator test-set annotations are reviewed at the case-structure level before quantitative evaluation.
Of \num{4597} reviewed case-structure pairs, \num{3002} are judged reliable and retained for the main evaluation; unreliable or inconclusive pairs are excluded without removing other structures of the same case.
To assess whether this exclusion influences the comparison between CADS and the baseline, board-certified radiologists manually correct 72 excluded case-structure pairs across nine representative structures, which are then re-evaluated for both models as a corrected mini-benchmark.
Per-structure exclusion records, error-type analyses, correction procedures, and mini-benchmark results are provided in Supplementary Note~5, and the review records and corrected labels are publicly released (see Data availability).

\subsubsection*{Independent evaluation cohorts}
\label{sec_04_methods_independent_eval_cohorts}
The University Hospital Zurich evaluation cohort comprises \num{2864} CT scans with oncological conditions commonly encountered in radiation oncology practice.
The distribution of primary disease sites is provided in Supplementary Table~24, with central nervous system malignancies, bone tumors, and head and neck cancers representing the largest groups.
Reference masks are either generated in MIM software and validated by physicians or contoured directly by radiation oncologists according to standard clinical protocols, and they serve as the reference standard for all quantitative metrics.
These annotations are produced during routine clinical care by clinicians independent of CADS-dataset curation, model development, and the TotalSegmentator annotation review.

To assess generalization beyond the public test sets, we additionally evaluate CADS on six independent external cohorts comprising \num{1160} CT scans, covering thoracic organs at risk, abdominal organs, pelvic bones, head-and-neck organs at risk, vertebrae, and cardiac substructures.
Dataset composition, per-cohort results, and interpretation of domain-specific performance patterns are provided in Supplementary Note~6.

\subsection*{Statistics and reproducibility}
\label{sec_04_methods_statistics_reproducibility}

No statistical method is used to predetermine sample size. 
All eligible CT scans from the public datasets and hospital cohorts described in the Methods section are included. 
Some data are excluded from quantitative evaluation only: reference annotations judged unreliable in the TotalSegmentator label review, structures with ground-truth volumes below 10\% of the corresponding median organ volume, and dataset-specific cases described in Supplementary Note~5. 
All exclusion criteria are predefined and applied identically to CADS-model and baseline predictions.

The experiments are not randomized because this is a retrospective study using previously acquired imaging data. 
Investigators are not blinded during annotation review or qualitative outcome assessment. 
Quantitative metrics are computed automatically against fixed reference annotations.

CADS-model and TotalSegmentator are compared separately for each structure using paired case-level scores. 
Two-sided paired $t$-tests or Wilcoxon signed-rank tests are used according to the distribution of paired differences assessed by the Shapiro–Wilk test. 
Statistical significance is defined as $p<0.05$, without correction for multiple testing. 
Confidence intervals (95\%) for mean scores are estimated by non-parametric percentile bootstrap resampling with \num{10000} iterations.
Statistical procedures for flavor selection are described in Supplementary Note~2, and further details on analyses and exclusions in Supplementary Note~5.

\subsection*{Data availability}
\label{dataav_methods}
The CADS-dataset, comprising paired CT volumes and curated whole-body annotations, is publicly available via Hugging Face at \url{https://huggingface.co/datasets/mrmrx/CADS-dataset}~\cite{xu2026cadsdataset}, subject to the licenses of the original source datasets.
The TotalSegmentator test-set review records and manually corrected labels are available via GitHub at \url{https://github.com/murong-xu/CADS}~\cite{xu2026cads}.
The University Hospital Zurich evaluation data are not publicly available, as they consist of protected patient imaging data obtained from routine clinical care.
Aggregated per-structure results for this cohort are provided in the Supplementary Information.
Source data are provided with this paper.

\subsection*{Code availability}
The CADS codebase is available via GitHub at \url{https://github.com/murong-xu/CADS}; the version used in this study is release v1.0.3~\cite{xu2026cads}.
The trained weights of the reference CADS-model are available as release v1.0.0 in the same repository~\cite{xu2026cads}, with the complementary model variants distributed as separate releases.
The 3D Slicer extension, which provides a graphical user interface for direct segmentation, is available together with its source code and installation instructions at \url{https://github.com/murong-xu/SlicerCADSWholeBodyCTSeg}~\cite{xu2026slicercads}.

\clearpage
\bibliographystyle{ieeetr}
\bibliography{bibliography}

\clearpage
\section*{Acknowledgements}
\label{acknowledgemnts_methods}
We thank the data providers and original authors of the public datasets integrated into the CADS-dataset.

SpineR software (bonescreen GmbH) was used to support the correction of vertebral reference segmentations in the curated evaluation subset.
Labelata GmbH contributed to the review and correction of segmentation labels, with particular contributions from Muhammad Tayyab Zia.
The three-dimensional visualizations presented in Figures~\ref{fig:results_whole_body_parts_visualization},~\ref{fig:results_overview_style_1},~\ref{fig:results_visualization}, and~\ref{fig:3dslicer} were created by the authors from the imaging and segmentation data analyzed in this study using 3D Slicer~\cite{fedorov20123d}.

\section*{Funding statement}
\label{funding_statement}
This research was supported by the Helmut Horten Foundation and the Comprehensive Cancer Center Zurich (C3Z Precision Oncology Funding Program, OMD-ZH project).
S.S. and B.M. acknowledge support from the Swiss National Science Foundation (grant 232553).
Funding was provided by the European Research Council (ERC) under the European Union's Horizon 2020 research and innovation programme (grant agreement No. 101045128, iBack-epic, ERC-2021-COG).
This work was supported by the German Research Foundation (DFG) through project 283653538, project WE 2709/3-3 (Computational bone health phenotyping in multiple myeloma patients), project WE 2709/3-1 (Whole body image analysis for diagnosing patients with monoclonal plasma cell disorders), and the Collaborative Research Center (CRC) 1444 (grant No. 427826188).
This work also received funding from the European Union's Seventh Framework Programme (FP7), ICT Call 8 (FP7-ICT-2011-8), under grant agreement No. 318068 (VISual Concept Extraction competition in RAdioLogy, VISCERAL).
Further support was provided by the Research Campus MODAL (German Federal Ministry of Research, Technology and Space [BMFTR], grant No. 3FO18501).

\section*{Author contributions}
\label{authorcont_methods}
M.X., T.A., F.Na., and B.M. designed the study. 
M.X., T.A., F.Na., I.E.H., and S.E. collected the data.
M.X., T.A., and F.Na. performed data analysis, model development, and model validation. 
M.X. drafted the manuscript, with input and revisions from T.A. and F.Na.
M.X., E.d.l.R., and J.D. contributed to figure preparation. 
M.X. developed the 3D Slicer plugin.
M.F., S.M.C., and S.T.L. contributed to the clinical evaluation.
I.E.H., S.E., and J.C.P. managed data anonymization and annotation.
I.E.H., S.E., M.-A.W., G.L., M.K.O., and J.S.K. contributed to the new data release.
S.S.B., G.B., J.H., F.Ne., and R.H. contributed to data processing.
S.S., B.W., and A.S. optimized label quality.
N.M. and S.K. reviewed label quality. 
R.G. and H.M. optimized model inference.
A.F., R.K., J.W., E.d.l.R., S.Z., and S.E.C. provided scientific advice. 
T.A. and B.M. supervised the study.
All authors reviewed and approved the manuscript.

\section*{Competing interests}
\label{competing_interests}
The authors declare no competing interests.

\clearpage
\clearpage 

\definecolor{headerdark}{gray}{0.27}
\definecolor{unshaded}{gray}{1.00}
\definecolor{shaded}{gray}{0.94}
\definecolor{totalband}{gray}{0.88}

\newcommand{\hdrcell}{\color{white}\bfseries}

\newcolumntype{C}[1]{>{\centering\arraybackslash}m{#1}}
\newcolumntype{L}[1]{>{\raggedright\arraybackslash}m{#1}}
\begin{table}[ht!]
\centering
\small 
\resizebox{\textwidth}{!}{
\begin{tabular}{
L{3.8cm}
C{1.2cm}
C{1.2cm}
C{1.2cm}
C{1.3cm}
C{1.5cm}
C{1.3cm}
C{2.8cm}
C{1.8cm}
L{5.5cm}
L{3.5cm}
}
\toprule
\rowcolor{headerdark}
\hdrcell Data Source & 
\hdrcell \#Vols &
\hdrcell \# Ann.\newline Vols &
\hdrcell \# Ann.\newline Struct. &
\hdrcell Body\newline Reg. &
\hdrcell Ann.\newline Method &
\hdrcell \#Cent. &
\hdrcell Countries &
\hdrcell Contrast &
\hdrcell Pathology &
\hdrcell License \\
\midrule
\rowcolor{shaded}
\textbf{VISCERAL Gold Corpus}~\cite{visceral}                 & 40       & 40           & 20          & WB, VR    & Human          & 3           & AT, DE, ES                 & Mix               & Pathologic: Bone Marrow Neoplasm, Lymphoma                        & Non-commercial (by agreement) \\
\rowcolor{unshaded}
\textbf{VISCERAL Gold Corpus-Extra}           & -        & 40           & 22          & WB, VR    & Human          & 3           & AT, DE, ES                 & Mix               & Pathologic: Bone Marrow Neoplasm, Lymphoma                        & Non-commercial (by agreement) \\
\rowcolor{shaded}
 \textbf{VISCERAL Silver Corpus}               & 127      & 127          & 20          & WB, VR    & Automated & 3           & AT, DE, ES                 & Mix               & Pathologic: Bone Marrow Neoplasm, Lymphoma                        & Non-commercial (by agreement) \\               
 \rowcolor{unshaded}                                                 
 \textbf{KiTS}~\cite{kits}                                 & 300      & 300          & 1           & AB        & Human          & 1           & US                         & Contrast-enhanced & Pathologic: Kidney Tumor                                          & CC BY-NC-SA 4.0                                                     \\
\rowcolor{shaded}
 \textbf{LiTS}~\cite{lits}                                 & 201      & 201          & 1           & AB        & Human          & 7           & DE, NL, CA, IL, FR         & Contrast-enhanced & Pathologic: Liver Tumor (Primary \& Metastasis)                    & CC BY-NC-SA 4.0                                                     \\
 \rowcolor{unshaded}
 \textbf{BTCV-Abdomen}~\cite{btcv}                         & 50       & 30           & 13          & AB        & Human          & 1           & US                         & Contrast-enhanced & Pathologic: Colorectal Cancer, Hernia                             & CC BY 4.0                                                           \\
\rowcolor{shaded}
 \textbf{BTCV-Cervix}                          & 50       & 30           & 3           & PR        & Human          & 1           & NL                         & -                 & Pathologic: Cervical Cancer                                       & CC BY 4.0                                                           \\
 \rowcolor{unshaded}
 \textbf{CHAOS}~\cite{chaos}                                & 40       & 20           & 1           & AB        & Human          & 1           & TR                         & Contrast-enhanced & Healthy                                                           & CC BY-NC-SA 4.0                                                     \\
\rowcolor{shaded}
 \textbf{AbdomenCT-1K}~\cite{abdomenct1k}                         & \num{1062}     & \num{1000}         & 4           & AB        & Hybrid         & 12          & DE, NL, CA, FR, IL, US, CN & Mix               & Pathologic: Liver, Pancreas, Kidney, Colon                        & CC BY 4.0                    \\
 \rowcolor{unshaded}
 \textbf{VerSe}~\cite{verse}                                & 374      & 374          & 22          & SP        & Hybrid         & -           & -                          & -                 & Pathologic: Spine (Fracture, Implants, Foreign Material)          & CC BY-SA 4.0                                                        \\
\rowcolor{shaded}
 \textbf{EXACT09}~\cite{exact09}                             & 40       & -            & -           & TH        & -              & 8           & -                          & Mix               & Mixed: Lung Disease (Healthy to Severe Pathologies)               & Restricted (challenge agreement)               \\
 \rowcolor{unshaded}
 \textbf{CAD-PE}~\cite{cadpe_pub}                               & 40       & -            & -           & TH        & -              & 6           & ES                         & Contrast-enhanced & Pathologic: Pulmonary Embolism                                    & CC BY 4.0                                                           \\
\rowcolor{shaded}
 \textbf{RibFrac}~\cite{ribfrac}                              & 660      & -            & -           & TH, AB    & -              & 1           & CN                         & -                 & Pathologic: Rib Fractures                                         & CC BY-NC 4.0                                                        \\
 \rowcolor{unshaded}
 \textbf{Learn2reg}~\cite{learn2reg}                            & 16       & 8            & 4           & AB        & Human          & -           & -                          & -                 & -                                                                 & Mixed: TCIA + CC BY 3.0                                             \\
\rowcolor{shaded}
 \textbf{LNDb}~\cite{lndb}                                 & 294      & -            & -           & TH        & -              & 1           & PT                         & -                 & Pathologic: Lung Nodules (Screening)                              & CC BY-NC-ND 4.0                                                          \\
 \rowcolor{unshaded}
 \textbf{LOLA11}~\cite{lola11}                               & 55       & -            & -           & TH        & -              & -           & -                          & -                 & Pathologic: Multiple Serious Abnormalities                        & Restricted (challenge agreement)     \\
\rowcolor{shaded}
 \textbf{SLIVER07}~\cite{sliver07}                             & 30       & 20           & 1           & TH        & Human          & Various     & -                          & Contrast-enhanced & Pathologic: Multiple Tumors, Cysts, Metastases                    & Restricted (challenge agreement)  \\
 \rowcolor{unshaded}
 \textbf{STOIC2021}~\cite{stoic}                            & \num{2000}     & -            & -           & TH        &     -    & 20          & FR                         & Mix               & Pathologic: COVID-19 Suspected                                    & CC BY-NC 4.0                                                        \\
\rowcolor{shaded}
 \textbf{CT-RATE}~\cite{hamamci2024developing}     & \num{3134}     & -            & -           & TH        & -              & 1           & TR                         & -                 & -                                                                 & CC BY-NC-SA 4.0 \\
 \rowcolor{unshaded}
 \textbf{EMPIRE10}~\cite{empire}                             & 60       & 60           & 1           & TH        & Hybrid         & Various     & NL, BE                     & -                 & Mixed: Lung Disease or Healthy                                    & Restricted (challenge agreement)     \\
\rowcolor{shaded}
 \textbf{AMOS}~\cite{amos}                                 & 200      & 200          & 15          & AB        & Hybrid         & 2           & CN                         & Mix               & Pathologic: Abdominal Tumors (Healthy Excluded)                   & CC BY 4.0                                                           \\
 \rowcolor{unshaded}
 \textbf{HaN-Seg}~\cite{hanseg}                              & 42       & 42           & 30          & HN        & Human          & 1           & SI                         & Contrast-enhanced                 & Pathologic: Mainly Head and Neck Tumors                                                                 & CC BY-NC-ND 4.0                                                     \\
\rowcolor{shaded}
 \textbf{HaN-Seg Extra Brain Labels}           & -        & 42            & 9           & HN        & Automated & 1           & SI                         & Contrast-enhanced                 & Pathologic: Mainly Head and Neck Tumors                                                                & CC BY-NC-ND 4.0                                                     \\
 \rowcolor{unshaded}
 \textbf{CT-ORG}~\cite{ctorg_data}                               & 140      & 140          & 5           & WB, AB    & Hybrid         & 8           & DE, NL, CA, IL, FR, US     & Mix               & Mixed: Liver Lesions (Benign \& Malignant), Bone \& Lung Metastasis & CC BY 3.0                                                           \\
 \arrayrulecolor{black!45}\midrule\arrayrulecolor{black}
\rowcolor{shaded}
 \textbf{LIDC-IDRI}~\cite{lidc_idri_data}                            & 997      & -            & -           & TH        & -              & 7           & -                          & -                 & Pathologic: Lung Nodules (Benign or Malignant)                    & CC BY 3.0                                                           \\
 \rowcolor{unshaded}
 \textbf{CT Lymph Nodes}~\cite{ctlymphnodes_data}                      & 174      & -            & -           & TH, AB    & -              & Various     & -                          & -                 & Pathologic: Lymphadenopathy (Non-cancer), Abdomen, Mediastinum    & CC BY 3.0                                                           \\
\rowcolor{shaded}
 \textbf{CPTAC-CCRCC}~\cite{cptac_ccrcc}                          & 258      & -            & -           & TH, AB    & -              & -           & -                          & -                 & Pathologic: Kidney Clear Cell Carcinoma                           & CC BY 3.0                                                           \\
 \rowcolor{unshaded}
 \textbf{CPTAC-LUAD}~\cite{cptac_luad}                           & 133      & -            & -           & TH, AB    & -              & -           & -                          & Mix               & Pathologic: Lung Adenocarcinoma                                   & CC BY 3.0                                                           \\
\rowcolor{shaded}
 \textbf{CT Images in COVID-19}~\cite{ct_covid19_data}                & 121      & -            & -           & TH        & -              & 4           & CN, JP, IT                 & -                 & Pathologic: COVID-19 Pneumonia                                    & CC BY 4.0                                                           \\
 \rowcolor{unshaded}
 \textbf{NSCLC Radiogenomics}~\cite{nsclc_radiomics_data}                  & 131      & -            & -           & TH        & -              & 2           & US                         & -                 & Pathologic: NSCLC (Non-Small Cell Lung Cancer)                    & CC BY 3.0                                                           \\
\rowcolor{shaded}
 \textbf{Pancreas-CT}~\cite{pancreasct_data}                          & 80       & -            & -           & AB        & -              & 1           & US                         & Contrast-enhanced & Healthy                                                           & CC BY 3.0                                                           \\
 \rowcolor{unshaded}
 \textbf{Pancreatic-CT-CBCT-SEG}~\cite{pancreatic_ctcbct_seg_data}               & 93       & -            & -           & TH, AB    & -              & 1           & US                         & Mix               & Pathologic: Pancreatic Cancer                                     & CC BY 4.0                                                           \\
\rowcolor{shaded}
 \textbf{RIDER Lung CT}~\cite{rider_lung_ct_data}                        & 59       & -            & -           & TH        & -              & 1           & US                         & Non-contrast      & Pathologic: NSCLC, Pulmonary Metastases                           & CC BY 4.0                                                           \\
 \rowcolor{unshaded}
 \textbf{TCGA-KICH}~\cite{tcga_kich}                           & 17       & -            & -           & AB        & -              & -           & -                          & -                 & Pathologic: Kidney Chromophobe Carcinoma                          & CC BY 3.0                                                           \\
\rowcolor{shaded}
 \textbf{TCGA-KIRC}~\cite{tcga_kirc}                           & 398      & -            & -           & AB        & -              & -           & -                          & -                 & Pathologic: Kidney Renal Clear Cell Carcinoma                     & CC BY 3.0                                                           \\
 \rowcolor{unshaded}
 \textbf{TCGA-KIRP}~\cite{tcga_kirp}                            & 19       & -            & -           & AB        & -              & -           & -                          & -                 & Pathologic: Kidney Papillary Cell Carcinoma                       & CC BY 3.0                                                           \\
\rowcolor{shaded}
 \textbf{TCGA-LIHC}~\cite{tcga_lihc}                           & 242      & -            & -           & AB        & -              & -           & -                          & -                 & Pathologic: Liver Hepatocellular Carcinoma (HCC)                  & CC BY 3.0                                                           \\
 \rowcolor{unshaded}
 \textbf{National Lung Screening Trial (NLST)}~\cite{nlst_data} & \num{7172}     & -            & -           & TH        & -              & 33          & US                         & Non-contrast      & -                                                                 & CC BY 4.0                                                           \\
 \arrayrulecolor{black!45}\midrule\arrayrulecolor{black}
\rowcolor{shaded}
 \textbf{TotalSegmentator}~\cite{totalsegmentator}                   & \num{1203}     & \num{1203}         & 104         & VR        & Hybrid         & 8           & CH                         & Contrast-enhanced & Mixed: 645 Pathologic (e.g., Tumor, Trauma), 404 Healthy           & CC BY 4.0                                                           \\
 \rowcolor{unshaded}
 \textbf{SAROS}~\cite{saros}                               & 900      & 900          & 11          & WB, VR    & Hybrid         & Various     & -                          & -                 & -                                                                 & Mixed: CC BY 3.0, CC BY 4.0, CC BY-NC 3.0 \\
 \arrayrulecolor{black!45}\midrule\arrayrulecolor{black}
\rowcolor{shaded}
 \textbf{New Hospital Data: CT-TRI}~\cite{ruhling2022automated}                              & 586      & -            & -           & AB         & -              & 1           & DE                          & Mix                 & Pathologic: Liver and Kidney Cancer                                                                 & CC BY-NC-SA 4.0\\
 \rowcolor{unshaded}
 \textbf{New Hospital Data: Head CT}         & 484      & -            & -           & HN        & -              & 1           & TR                         & -                 & -                                                                 & CC BY 4.0                                                                 \\
 \midrule
\rowcolor{totalband}
 \textbf{Total}                                & \num{22022}    & \num{4695}         & 167       &   -     &      -       &  $>100$        &       16                  &        Mix        &       Healthy and Pathologic                                                          &      -                                                            \\
\bottomrule
\end{tabular}
}
\caption{Summary of the CT collection in the CADS-dataset.
The \num{22022} volumes come from 40 sources in 16 countries, grouped (separated by horizontal rules) into public challenge datasets~\cite{grandchallenge}, collections from The Cancer Imaging Archive (TCIA)~\cite{tcia}, whole-body datasets, and newly collected hospital scans; \num{4695} volumes (21\%) contain partial annotations.
\#Vols, number of volumes; \#Ann. Vols, number of annotated volumes; \#Ann. Struct., number of annotated structures; Body Reg., body region (WB, whole body; HN, head and neck; TH, thorax; AB, abdomen; PR, pelvic region; SP, spine; VR, various regions); Ann. Method, annotation method (Human, manual; Automated, algorithm-generated; Hybrid, algorithm-generated and manually refined); \#Cent., number of acquisition centers (Various, multiple centers with the exact number not reported); Countries (AT, Austria; BE, Belgium; CA, Canada; CH, Switzerland; CN, China; DE, Germany; ES, Spain; FR, France; IL, Israel; IT, Italy; JP, Japan; NL, Netherlands; PT, Portugal; SI, Slovenia; TR, Turkey; US, United States); Contrast (Mix, both contrast-enhanced and non-contrast scans).
A dash (-) indicates information that is not available or not applicable.
}
\label{table:dataset_collection}
\end{table}

\clearpage 

\setlength{\tabcolsep}{1.8pt}
\renewcommand{\arraystretch}{0.98}
\newcolumntype{S}{>{\raggedright\arraybackslash}p{2.9cm}}
\newcolumntype{N}{>{\raggedleft\arraybackslash}p{1.0cm}}
\newcolumntype{V}{>{\raggedleft\arraybackslash}p{1.2cm}}
\newcommand{\target}[3]{#1 & \num{#2} & \num{#3}}

\begin{table}
\centering
\resizebox{\textwidth}{!}{%
\scriptsize
\begin{tabular}{@{}S N V @{\qquad} S N V @{\qquad} S N V @{\qquad} S N V@{}}
\toprule
\textbf{Structure} & \textbf{Cases} & \multicolumn{1}{c}{\textbf{Median size}} &
\textbf{Structure} & \textbf{Cases} & \multicolumn{1}{c}{\textbf{Median size}} &
\textbf{Structure} & \textbf{Cases} & \multicolumn{1}{c}{\textbf{Median size}} &
\textbf{Structure} & \textbf{Cases} & \multicolumn{1}{c}{\textbf{Median size}} \\
\midrule
\target{Spleen}{21097}{59202} &
\target{Kidney R}{18482}{22472} &
\target{Kidney L}{20110}{24004} &
\target{Gallbladder}{15429}{3633} \\
\target{Liver}{21191}{430341} &
\target{Stomach}{21118}{72764} &
\target{Aorta}{21342}{58888} &
\target{Inferior vena cava}{21046}{13554} \\
\target{Portal and splenic vein}{20602}{3603} &
\target{Pancreas}{20641}{17155} &
\target{Adrenal gland R}{20652}{960} &
\target{Adrenal gland L}{20406}{1114} \\
\target{Upper lobe of lung L}{21011}{307385} &
\target{Lower lobe of lung L}{21137}{275330} &
\target{Upper lobe of lung R}{18241}{281021} &
\target{Middle lobe lung R}{20725}{109989} \\
\target{Lower lobe of lung R}{21379}{304069} &
\target{Vertebra L5}{5413}{17458} &
\target{Vertebra L4}{6519}{17874} &
\target{Vertebra L3}{8733}{16921} \\
\target{Vertebra L2}{15049}{13274} &
\target{Vertebra L1}{19730}{13718} &
\target{Vertebra T12}{20811}{13119} &
\target{Vertebra T11}{20891}{12054} \\
\target{Vertebra T10}{20732}{11152} &
\target{Vertebra T9}{20352}{9930} &
\target{Vertebra T8}{19667}{8931} &
\target{Vertebra T7}{18750}{8357} \\
\target{Vertebra T6}{17932}{7649} &
\target{Vertebra T5}{17475}{7195} &
\target{Vertebra T4}{17332}{6738} &
\target{Vertebra T3}{17318}{6426} \\
\target{Vertebra T2}{17289}{6650} &
\target{Vertebra T1}{17255}{6165} &
\target{Vertebra C7}{16744}{2601} &
\target{Vertebra C6}{9792}{570} \\
\target{Vertebra C5}{3255}{499} &
\target{Vertebra C4}{1167}{3111} &
\target{Vertebra C3}{905}{3349} &
\target{Vertebra C2}{1276}{4007} \\
\target{Vertebra C1}{1249}{3317} &
\target{Esophagus}{21030}{9350} &
\target{Trachea}{17547}{9773} &
\target{Myocardium}{20619}{31054} \\
\target{Heart atrium L}{20352}{17734} &
\target{Heart ventricle L}{20574}{29043} &
\target{Heart atrium R}{20516}{22501} &
\target{Heart ventricle R}{20684}{38916} \\
\target{Pulmonary artery}{17701}{18132} &
\target{Brain}{1268}{358424} &
\target{Iliac artery L}{6803}{2701} &
\target{Iliac artery R}{6240}{3160} \\
\target{Iliac vein L}{5703}{6974} &
\target{Iliac vein R}{5901}{5545} &
\target{Small intestine}{17026}{36292} &
\target{Duodenum}{17601}{6784} \\
\target{Colon}{20347}{54292} &
\target{Urinary bladder}{4022}{41437} &
\target{Face}{4793}{47} &
\target{Humerus L}{16344}{9753} \\
\target{Humerus R}{16629}{10476} &
\target{Scapula L}{18569}{28550} &
\target{Scapula R}{18568}{27772} &
\target{Clavicle L}{17302}{6226} \\
\target{Clavicle R}{17310}{6731} &
\target{Femur L}{3628}{48277} &
\target{Femur R}{3573}{48168} &
\target{Hip L}{5290}{96869} \\
\target{Hip R}{5272}{97104} &
\target{Sacrum}{4494}{62408} &
\target{Gluteus maximus L}{4122}{131518} &
\target{Gluteus maximus R}{4188}{133948} \\
\target{Gluteus medius L}{4776}{63677} &
\target{Gluteus medius R}{4803}{62294} &
\target{Gluteus minimus L}{3784}{16844} &
\target{Gluteus minimus R}{3810}{17984} \\
\target{Deep muscle of back L}{21344}{86627} &
\target{Deep muscle of back R}{21343}{85689} &
\target{Iliopsoas muscle L}{16617}{3361} &
\target{Iliopsoas muscle R}{17524}{2838} \\
\target{Rib 1 L}{17318}{2850} &
\target{Rib 2 L}{17363}{3374} &
\target{Rib 3 L}{17483}{3992} &
\target{Rib 4 L}{18180}{4910} \\
\target{Rib 5 L}{19522}{5405} &
\target{Rib 6 L}{20547}{5947} &
\target{Rib 7 L}{20776}{6218} &
\target{Rib 8 L}{20870}{5471} \\
\target{Rib 9 L}{20893}{4948} &
\target{Rib 10 L}{20897}{3963} &
\target{Rib 11 L}{20862}{2313} &
\target{Rib 12 L}{19871}{838} \\
\target{Rib 1 R}{17316}{2883} &
\target{Rib 2 R}{17369}{3394} &
\target{Rib 3 R}{17487}{4034} &
\target{Rib 4 R}{18074}{5122} \\
\target{Rib 5 R}{19445}{5661} &
\target{Rib 6 R}{20488}{6046} &
\target{Rib 7 R}{20720}{6371} &
\target{Rib 8 R}{20825}{5648} \\
\target{Rib 9 R}{20856}{5078} &
\target{Rib 10 R}{20879}{3813} &
\target{Rib 11 R}{20848}{2206} &
\target{Rib 12 R}{19790}{759} \\
\target{Spinal canal}{21939}{16463} &
\target{Larynx}{16531}{751} &
\target{Heart}{20805}{137449} &
\target{Bowel space}{21367}{186750} \\
\target{Sigmoid colon}{6263}{13524} &
\target{Rectum}{5850}{9185} &
\target{Prostate}{3618}{6225} &
\target{Seminal vesicle}{3619}{2145} \\
\target{Mammary gland L}{10784}{94565} &
\target{Mammary gland R}{9602}{119226} &
\target{Sternum}{20786}{15512} &
\target{Psoas major muscle R}{20251}{2601} \\
\target{Psoas major muscle L}{20218}{2999} &
\target{Rectus abdominis R}{21059}{13604} &
\target{Rectus abdominis L}{21054}{14002} &
\target{White matter}{3336}{1106} \\
\target{Gray matter}{1729}{163903} &
\target{CSF}{2612}{1546} &
\target{Scalp}{4622}{12311} &
\target{Eyeball}{1071}{4283} \\
\target{Compact bone}{4952}{861} &
\target{Spongy bone}{3578}{892} &
\target{Vascular structure of head}{2319}{2125} &
\target{Rectus eye muscle}{1083}{503} \\
\target{Carotid artery L}{17746}{510} &
\target{Carotid artery R}{16281}{175} &
\target{Arytenoid cartilage}{4522}{60} &
\target{Mandible}{5080}{2217} \\
\target{Brainstem}{1234}{5551} &
\target{Buccal mucosa}{1108}{761} &
\target{Oral cavity}{2244}{4116} &
\target{Cochlea L}{1068}{33} \\
\target{Cochlea R}{1104}{39} &
\target{Cricopharyngeus}{10050}{508} &
\target{Cervical esophagus}{16216}{549} &
\target{Anterior eyeball L}{403}{239} \\
\target{Anterior eyeball R}{397}{240} &
\target{Posterior eyeball L}{1035}{731} &
\target{Posterior eyeball R}{945}{457} &
\target{Lacrimal gland L}{771}{50} \\
\target{Lacrimal gland R}{711}{57} &
\target{Submandibular gland L}{1350}{904} &
\target{Submandibular gland R}{1418}{782} &
\target{Thyroid}{17074}{3424} \\
\target{Glottis}{7573}{566} &
\target{Supraglottis}{3951}{1138} &
\target{Lips}{1740}{2231} &
\target{Optic chiasm}{1019}{72} \\
\target{Optic nerve L}{1030}{152} &
\target{Optic nerve R}{1031}{157} &
\target{Parotid gland L}{1301}{4629} &
\target{Parotid gland R}{1313}{4738} \\
\target{Pituitary gland}{1045}{96} &
\target{Subcutaneous tissue}{22022}{1410735} &
\target{Muscle}{22022}{1609604} &
\target{Abdominopelvic cavity}{21990}{1280645} \\
\target{Thoracic cavity}{22001}{1329453} &
\target{Bones}{22022}{590596} &
\target{Gland structure}{18294}{2879} &
\target{Pericardium}{21283}{205442} \\
\target{Breast implant}{3341}{75} &
\target{Mediastinum}{21419}{175396} &
\target{Spinal cord}{21703}{17388} &
& & \\
\bottomrule
\end{tabular}
}
\caption{Summary statistics for the 167 segmentation targets in the CADS-dataset.
Cases, number of the \num{22022} CT volumes in which the structure is present; Median size, median number of voxels occupied by the structure across those volumes, computed after resampling to \SI{1.5}{mm} isotropic spacing (one voxel corresponds to \SI{3.375}{mm^3}); L, left; R, right.}
\label{table:cads-dataset_statistics}
\end{table}


\begin{figure*}[!htb]
    \centering
    \includegraphics[trim=0mm 0mm 0mm 0mm, clip, width=\textwidth]{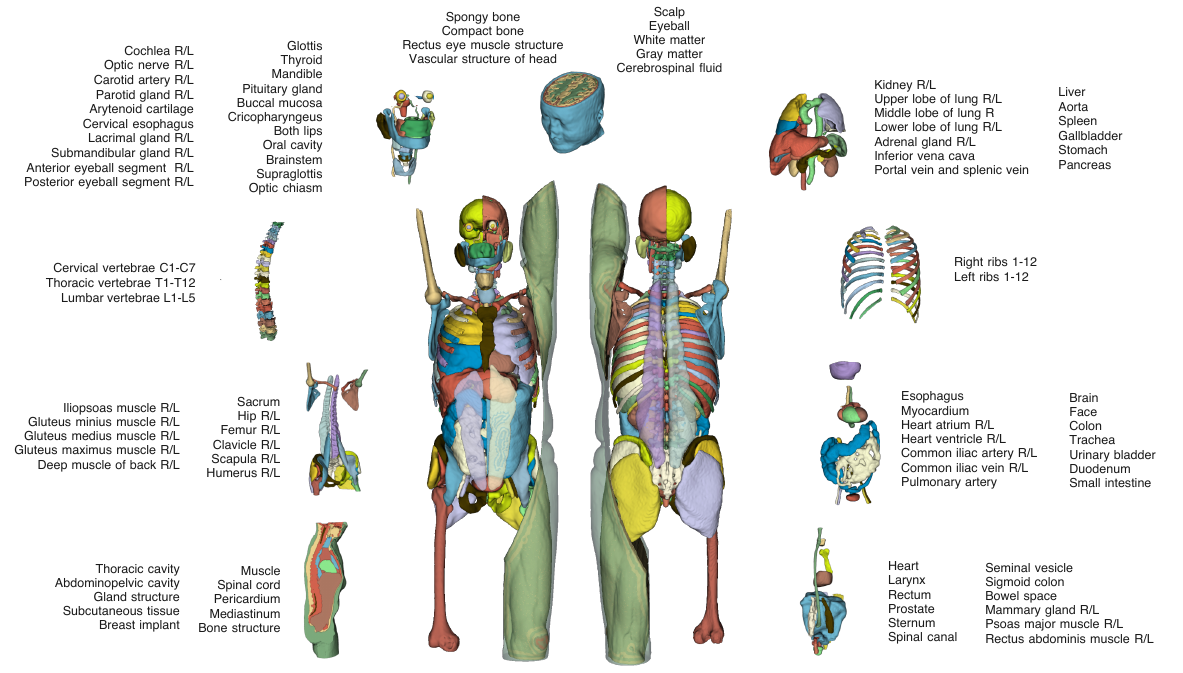}
    \caption{Anatomical coverage of CADS.
    CADS segments 167 structures from head to knees.
    The center shows anterior and posterior three-dimensional renderings of a whole-body segmentation, and the surrounding panels show the nine model groups.
    Colors distinguish individual structures only. R, right; L, left.}
    \label{fig:results_whole_body_parts_visualization}
\end{figure*}

\clearpage 

\clearpage 

\clearpage 

\clearpage 

\begin{figure*}[!htb]
    \centering
    \includegraphics[trim=0mm 0mm 0mm 0mm, clip, width=\textwidth]{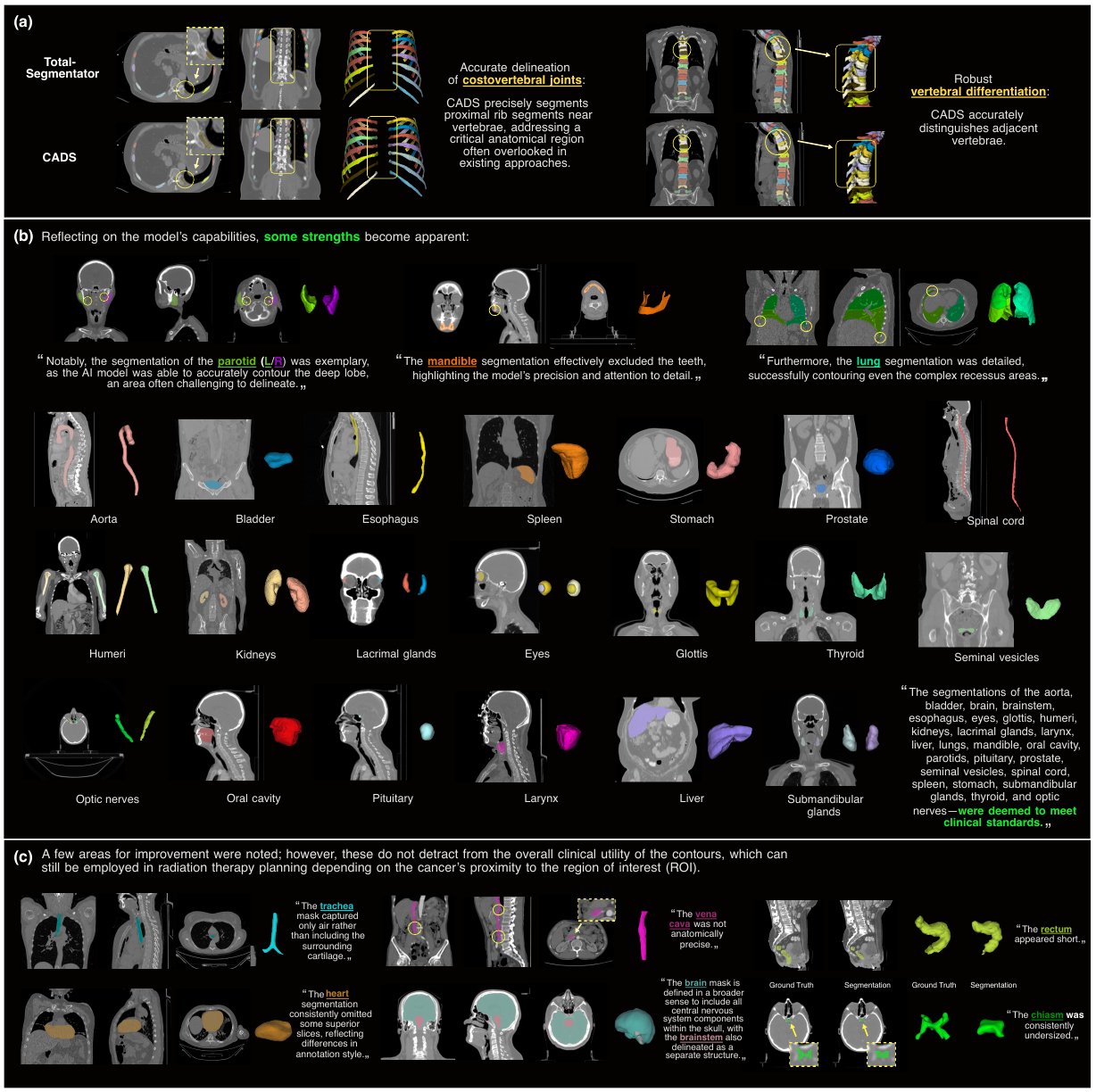}
    \caption{Qualitative assessment of CADS-model segmentations.
    (a) Comparison of skeletal segmentations from existing methods and the CADS-model. CADS-model segmentations better follow standard anatomical definitions, with clearer delineation of the costovertebral joints and separation of adjacent vertebrae.
    (b, c) Examples from the University Hospital Zurich cohort reviewed by a radiation oncologist. Three cases around the median Dice score were reviewed for each structure; representative cases are shown with the reviewer’s comments.
    (b) Structures meeting clinical standards in the reviewed cases.
    (c) Structures identified for refinement or showing differences in annotation definitions.
    Colors distinguish individual structures. L, left; R, right; ROI, region of interest.}
    \label{fig:results_visualization}
\end{figure*}

\clearpage 

\end{document}